\documentclass[letterpaper,12pt]{article}  

\usepackage{amsmath}
\usepackage{amssymb}
\usepackage[USenglish]{babel}
\usepackage[utf8]{inputenc}
\usepackage[T1]{fontenc}
\usepackage{color}
\usepackage{graphicx}
\usepackage{enumerate}
\usepackage{ulem}
\usepackage{dcolumn}
\usepackage{multirow}
\usepackage{url}
\usepackage{authblk}

\usepackage{colortbl}

\newlength{\textlarg}

\definecolor{darkred}{rgb}{.8,.0,.0}
\definecolor{gris}{rgb}{0.75,0.75,0.75}

\graphicspath{{images/}}

\begin{document}

\title{Predicting links in ego-networks using temporal information}

\author[1]{Lionel Tabourier\thanks{corresponding author: \texttt{lionel.tabourier@ens-lyon.org}}}

\author[2]{Anne-Sophie Libert}
\author[2]{Renaud Lambiotte}

\affil[1]{Sorbonne Universités, UPMC Université Paris 06, CNRS\\ LIP6 UMR 7606, 4 place Jussieu 75005 Paris, France}
\affil[2]{naXys, University of Namur\\ Rempart de la Vierge 8, 5000 Namur, Belgium}

\date{}

\maketitle

\begin{abstract}

Link prediction appears as a central problem of network science, as it calls for unfolding the mechanisms that govern the micro-dynamics of the network.
In this work, we are interested in \textit{ego-networks}, that is the mere information of interactions of a node to its neighbors, in the context of social relationships.
As the structural information is very poor, we rely on another source of information to predict links among egos' neighbors: the timing of interactions.
We define several features to capture different kinds of temporal information and apply machine learning methods to combine these various features and improve the quality of the prediction. 
We demonstrate the efficiency of this temporal approach on a cellphone interaction dataset, pointing out features which prove themselves to perform well in this context, in particular the temporal profile of interactions and elapsed time between contacts.

\end{abstract}


\section{Introduction}

In  recent years, networks have become a ubiquitous way of representing any kind of interacting systems ranging from metabolic protein interactions to online social networks.
This trend is justified by the simplicity of the representation, combined with the technical possibility of storing and processing large-scale datasets.
In most cases though, the observer only has a partial view of the network, and achieving a comprehensive mapping of the interactions is often a challenging task.
Big data collection campaigns have been set in various fields, notably biological networks, or Internet mapping, but collecting large amounts of data remains expensive in both space and time.
%
%
In addition to that cost, metrological problems may bias the crawling process and compromise the reliability of the data.
%
%
When it comes to social data, the problem often originates in the traditional data collection methods, which are not suited for large-scale analysis, such as individual surveys.
Online social networks allow to access larger datasets, however, data providers often restrict the access to their resources for commercial, technical or legal reasons. 
Similarly, even private companies, for instance mobile phone operators, have a restricted view of a social system, as they only have full information about their clients and are blind to the connections between clients of other companies.


Analyzing local structures in networks consequently appears as a possible way to circumvent these issues.
In sociology, ego-centered networks have been  studied for a long time~\cite{freeman1982centered} and measures have been proposed to describe and understand the local structural environments around specific nodes~\cite{everett2005ego,stoica2009structure}.
More recently, the question of how to adequately define the notion of community in this context has been an important focus of interest~\cite{friggeri2011triangles,mcauley2012learning,danisch2012towards}.
In this work, we consider the following problem: knowing the interactions of a node with its direct neighbors, can we guess if there are existing links between these neighbors?
In other words, \textit{``among someone's friends, who are likely to know each other?''}
This is a typical link prediction problem, but in this case structural information about the network is lacking.
Hence, we resort to other sources namely temporal information, to discover links between nodes of a social network.

The link prediction problem in networks is often formulated as inferring which links may appear or not in the future from the observed structure of the network, see for example~\cite{liben2007link}.
This can be formulated as a machine learning task using learning features, which are related to the probability for a node to appear.
Structural features are often used to that purpose, for example the number of common neighbors, hitting time etc. 
There are many available metrics which can be found in surveys~\cite{lu2011link}.
Other kinds of features are also available, such as node-level attributes~\cite{bliss2014evolutionary}, or interaction-level attributes~\cite{merritt2013detecting}.
When considering link prediction in social networks, one should mention the class imbalance problem: a sparse network implies the fact that there are much more pairs of nodes than actual links.
It implies that there is a high risk of misclassification by increasing the number of predictions.
Efforts have been made to alleviate this acute problem, in particular, by using supervised learning techniques that allow to group pairs of nodes in categories for link prediction and, therefore, reduce the imbalance effect~\cite{lichtenwalter2010new}.


Interaction dynamics is also a valuable source of information.
For example, it is known that the pace and length of communications give clues about the type of relationship involved: family, commercial, friendship, etc \cite{eagle2009inferring}.
Several works exploited this for link prediction-related purposes using pattern frequencies to infer which interactions are most likely in the near future~\cite{tylenda2009towards,bringmann2010learning}, or predicting link decay from the measure of the elapsed time since the last interaction~\cite{raeder2011predictors}. 
In other contexts, temporal information was also incorporated in order to predict transitions between venues in cities~\cite{noulas2015topological}.
In this work, our goal is to extract information from the interaction dynamics to reveal existing links in ego-centered social networks.
Considering a phone call dataset, where a link represents the existence of a social interaction between two users, the scenario is that we only have local information on the interaction network of specific nodes.
It is then a minimal version of the ego-network, as it involves the node and its direct neighbors\footnote{Note that in other contexts, some authors refer to the ego-network as the links of an ego to its neighbors and the connections among them.}.
There is very little structural information available and hence, we use temporal information to rank pairs among the neighbors of an ego node.
A high-ranked pair should feature nodes of the same social circle, which are prone to interact with each other.
We also aim at point out temporal features, which are particularly informative in predicting links.


We design several types of features from the timing of interactions.
Then we tackle the problem as a ranking combination issue.
Each feature provides us with a ranking, which indicates pairs of neighbors  likely to be connected.
Following a strategy similar to~\cite{pujari2012supervised}, we combine these rankings in a supervised framework to draw as much information as possible from these features, so that the resulting ranking should rank high the pairs which are most likely to be connected.
We first use traditional classification methods to do so, as given in~\cite{liben2007link} or \cite{lichtenwalter2010new}, and show their limits as the number of predictions cannot be set according to our needs.
For this purpose, we use the learning-to-rank framework in~\cite{tabourier2014rankmerging}, especially designed for link prediction in large networks.
The benefit of using learning-to-rank instead of classification methods is that we predict exactly $T$ links by considering the top-$T$ pairs of our ranking.


We describe in Section~\ref{sec:data} the phone call and text messages dataset under examination in this article.
Then, in Section~\ref{sec:prediction_timing} we expose how the temporality of interactions can be used for predicting links in such datasets.
After describing the protocol of evaluation and the static benchmark that will be used for comparison, we propose temporal features which aim at guessing links among the neighbors of ego nodes.
We explain how these features are used in order to obtain rankings, where highly-ranked pairs are more likely to be connected.
In Section~\ref{sec:combin}, we propose supervised strategies to combine these rankings in order to obtain the best possible predictions, classification, as well as learning to rank techniques.

\section{Dataset}
\label{sec:data}

\subsection{Preprocessing}

The dataset under examination is a collection of communications made among a subset of anonymized subscribers to a european cellphone service provider.
It contains around $ 14.3 \cdot 10^6 $ calls and $ 28.8 \cdot 10^6 $ text messages made between any pair of users in the dataset during a one-month period.
Henceforth, we make the distinction between calls and text messages, because we assume that these means of communication are not used for the same purposes by the same people.
Calls can be represented as a list of quadruplets, $\left\lbrace source, destination, timestamp, duration \right\rbrace$.
Calls with null duration, corresponding to unanswered phone calls, have been filtered out of the dataset.
Text messages are stored as triplets, $\left\lbrace source, destination, timestamp \right\rbrace$.

The usual network representation of such data consists in describing users as nodes and the existence of at least one interaction between two users as a link. 
These links may be assigned a certain direction depending on who is calling/texting whom.
The total number of interactions (either calls or messages) between two nodes $i$ and $j$ during the whole record period will be referred to as the \textit{weight}, $ w(i,j)$, of this link.

As we are interested in the social groups underlying the communication network, we filter out calls and text messages which are not indicative of a lasting social relationship. 
We only consider calls on bidirectional links, that is to say links which have been activated in both directions \cite{onnela2007structure}.
Except for this step, interactions between users are considered as undirected.
The data comes down to 1,241,865 nodes, and 1,514,490 links  -- indifferently call or message links --
corresponding to 10,934,277 phonecalls and 27,060,340 text messages after preprocessing.

\begin{figure}[h!]
\begin{center}
\includegraphics[angle=-90,width=0.45\linewidth]{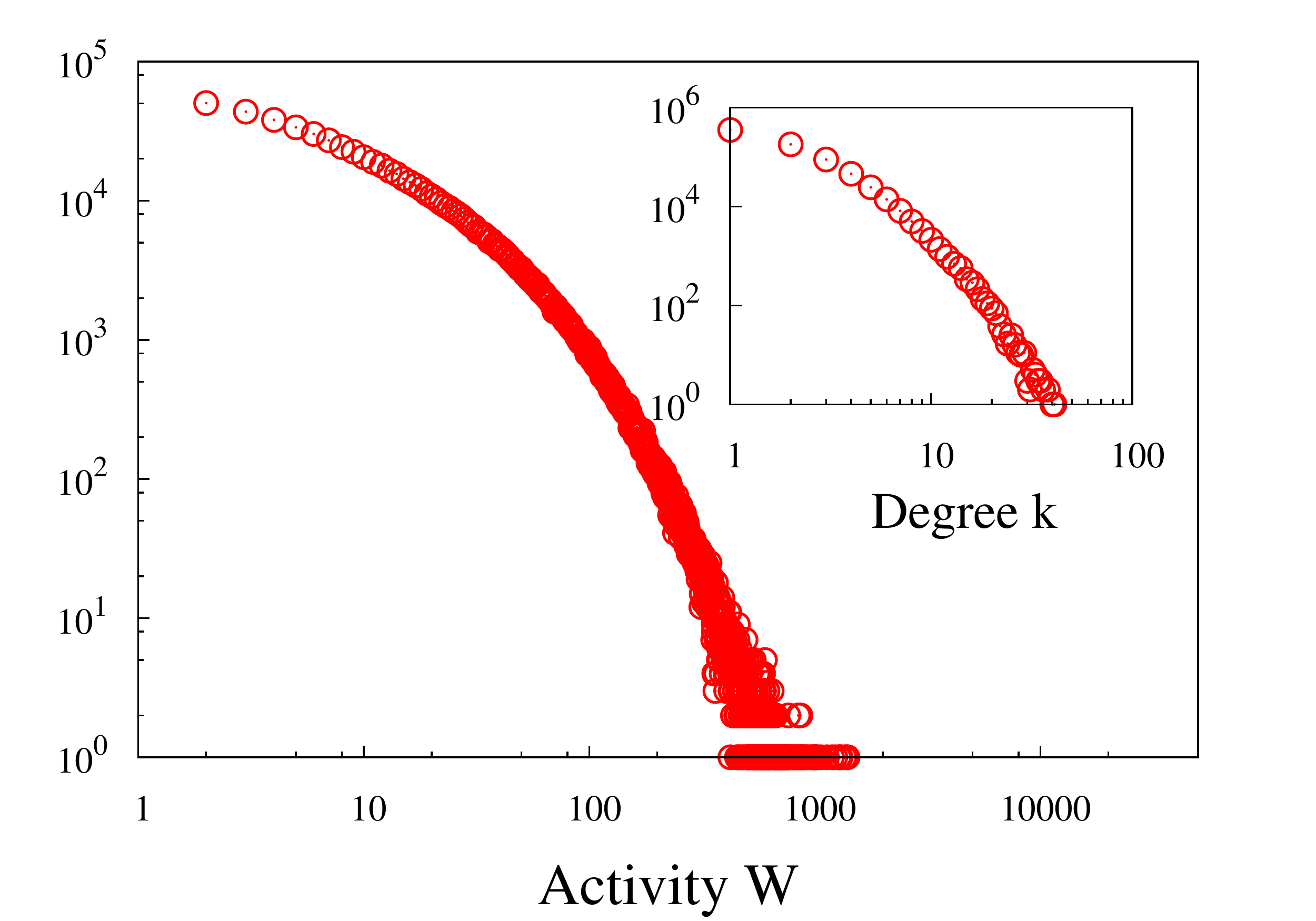}
\includegraphics[angle=-90,width=0.45\linewidth]{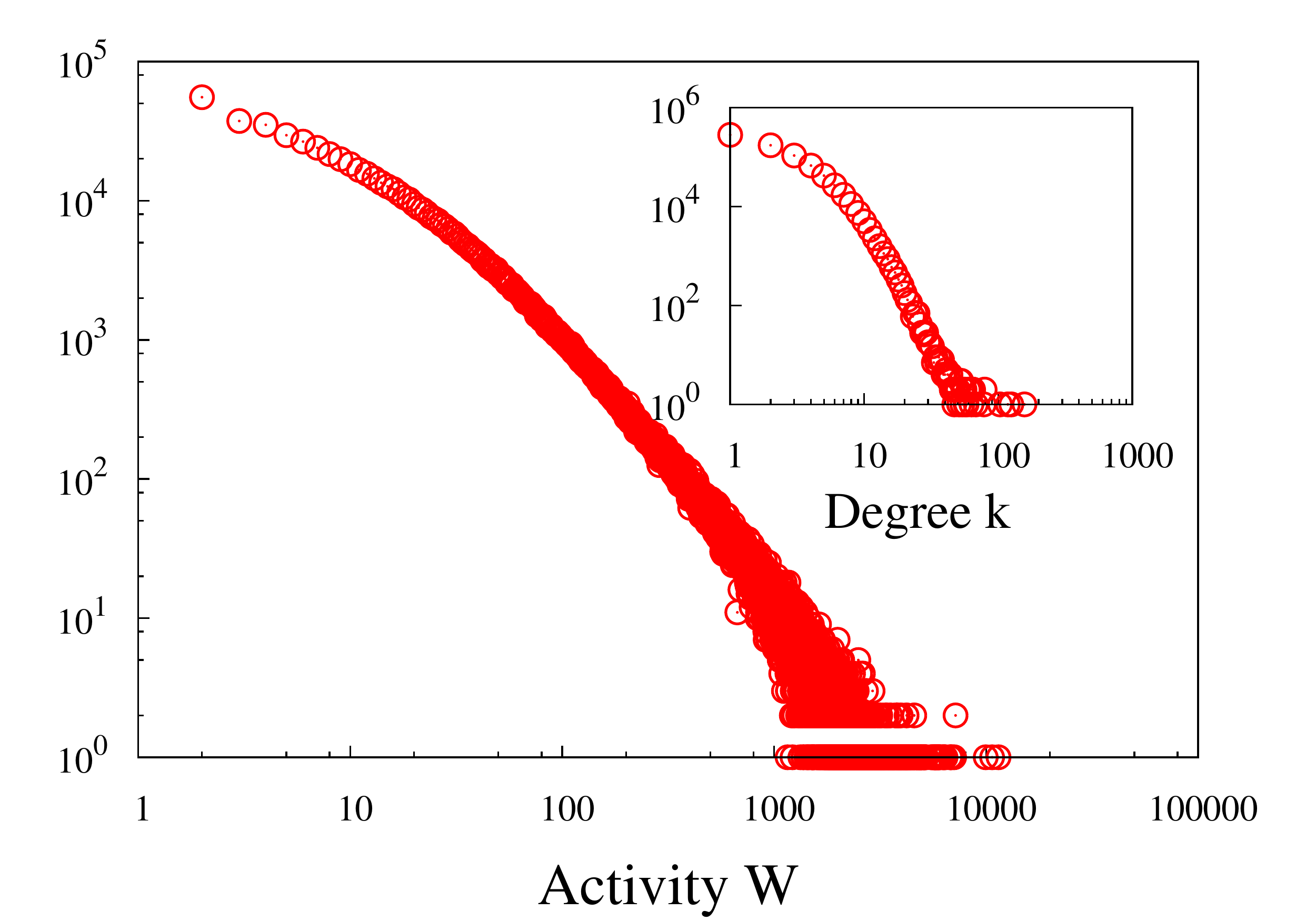}
\end{center}
\caption{\label{fig:distribs}
Activity and degree (inset) distributions in the dataset, for both  phonecalls (left) and text messaging (right) networks.}
\end{figure}

From now on, the network is regarded as a set of isolated \textit{ego-networks}, that is to say the interactions between a central node and its direct neighbors.
Nodes have heterogeneously distributed degrees and weights regarding both phonecalls and text messages, see Figure~\ref{fig:distribs}.
%
%
It is known that the prediction quality depends on the degree of the central node as underlined in~\cite{comar2011linkboost}.
Typically it is less efficient on low degree nodes because of the lack of information.
We, therefore, group nodes together into degree classes.
The learning process will be made on each of these sets separately to improve performances.


\subsection{Ego-networks specificities}


We consider a scenario where the only information available is the timing (and duration for calls) of interactions of a node to its neighbors, the information about the network structure is poor.
The temporal patterns of these interactions bear the trace of underlying social circles, and as such they enable us to predict the links existing in the neighborhood of the ego node.
Former works have stressed the dramatic effect of class imbalance on link prediction problems in social networks, especially in mobile phone networks~(\cite{lichtenwalter2010new,comar2011linkboost}).
The fact that there are much more pairs of nodes than links in the network makes the prediction and its evaluation tricky.
The typical order of magnitude of the classes ratio for a network of $N$ nodes is $ O(1/N)$.
However, in case of ego-networks, the class-imbalance effect is less of a problem, since the neighbors of a degree $k$ node have at most $ k(k-1)/2 $ links among themselves.
%
%
A direct consequence of the lack of structural information present in ego-networks is that standard algorithms, for instance based on common neighbors, are unable to predict links between two nodes better than purely random predictions.

\section{Prediction based on temporal information}\label{sec:prediction_timing}

In this section, we present the protocol used to evaluate how the temporal information improves the quality of link prediction among the neighbors of an ego node.
For this purpose, we define metrics that allow to rank pairs of nodes, where the highest ranked pairs are the most likely to be connected.

\subsection{Protocol and prediction evaluation}

For each degree class $k$, that is the degree of the ego node, we divide ego-networks in three sets according to the following proportions: learning set (60\%), validation set (20\%) and test set (20\%).
If there are $N$ egos in a set, we rank the $ N \cdot k \cdot (k -1) /2 $ pairs of neighbors in the union of the ego-networks.
The presence or absence of a link  between two neighbors in the learning set is supposed to be known and will be used during the learning phase of the protocol, while the performance of the whole procedure is evaluated on the test set.
The validation set will be used to fix the parameters of the prediction method as discussed later.

The process is then divided into two parts, an unsupervised ranking part followed by a supervised aggregation of rankings.
During the first part, pairs of nodes are ranked according to a metric $m$.
$m$ is chosen to be correlated with the probability of existence of a link between neighbors.
We also use consensus-based strategies to obtain rankings combined from the metric-based rankings.
The quality of the various rankings produced is assessed by measuring the numbers of true and false positive predictions on the top pairs and usual related quantities, namely precision ($Pr$), recall ($Rc$) and F-score.
Let us remind that the F-score is defined as $ \frac{2 \cdot Pr \cdot Rc}{Pr + Rc}$.
In the line of~\cite{davis2006relationship}, we use precision-recall curves to visualize the performances of the prediction.
We also plot F-scores as a function of the number of predictions, as this quantity is proportional to the number of true positive for a given number of predictions.
Then we mix the rankings following supervised learning methods to obtain a prediction as accurate as possible on the various degree classes.

\subsection{Static benchmarks \label{sec:bench}}

The quality evaluation is made by comparison to benchmarks, which rely on the basic structural information.
For the comparison to be as fair as possible, we test a few ranking metrics and keep the most efficient one.
Each pair of neighbors $ (i,j) $ of ego $e$, with degree $k(e)$ and total weight $W(e) = \Sigma _i w(e,i) $, is given a score $ s(i,j) $ depending on the weights $ w(e,i)$ and $ w(e,j)$, which is the only structural information available here\footnote{Note that by convention, the pair $ (i,j) $ of neighbors of ego $e$ is considered distinct from the pair $ (i,j) $ of neighbors of $e'$. Hence, there are duplicates among the ranked pairs which may predicted twice, but this event is rare as it concerns less than $1$ pair over $1000$ and have practically no impact on the prediction.}. 
%
The static benchmark metrics are:
\begin{itemize}
\item[$ \bullet $] $s_1(i,j) = w(e,i) \cdot w(e,j)$
\item[$ \bullet $] $s_2(i,j) = w(e,i) + w(e,j)$
\item[$ \bullet $] $s_3(i,j) = max (w(e,i) , w(e,j))$ 
\item[$ \bullet $] $s_4(i,j) = w(e,i) \cdot w(e,j) / k(e) $
\item[$ \bullet $] $s_5(i,j) = w(e,i) \cdot w(e,j) / W(e) $ 

\end{itemize}

\begin{figure}[h!]
\begin{center}
\includegraphics[angle=-90,width=0.6\linewidth]{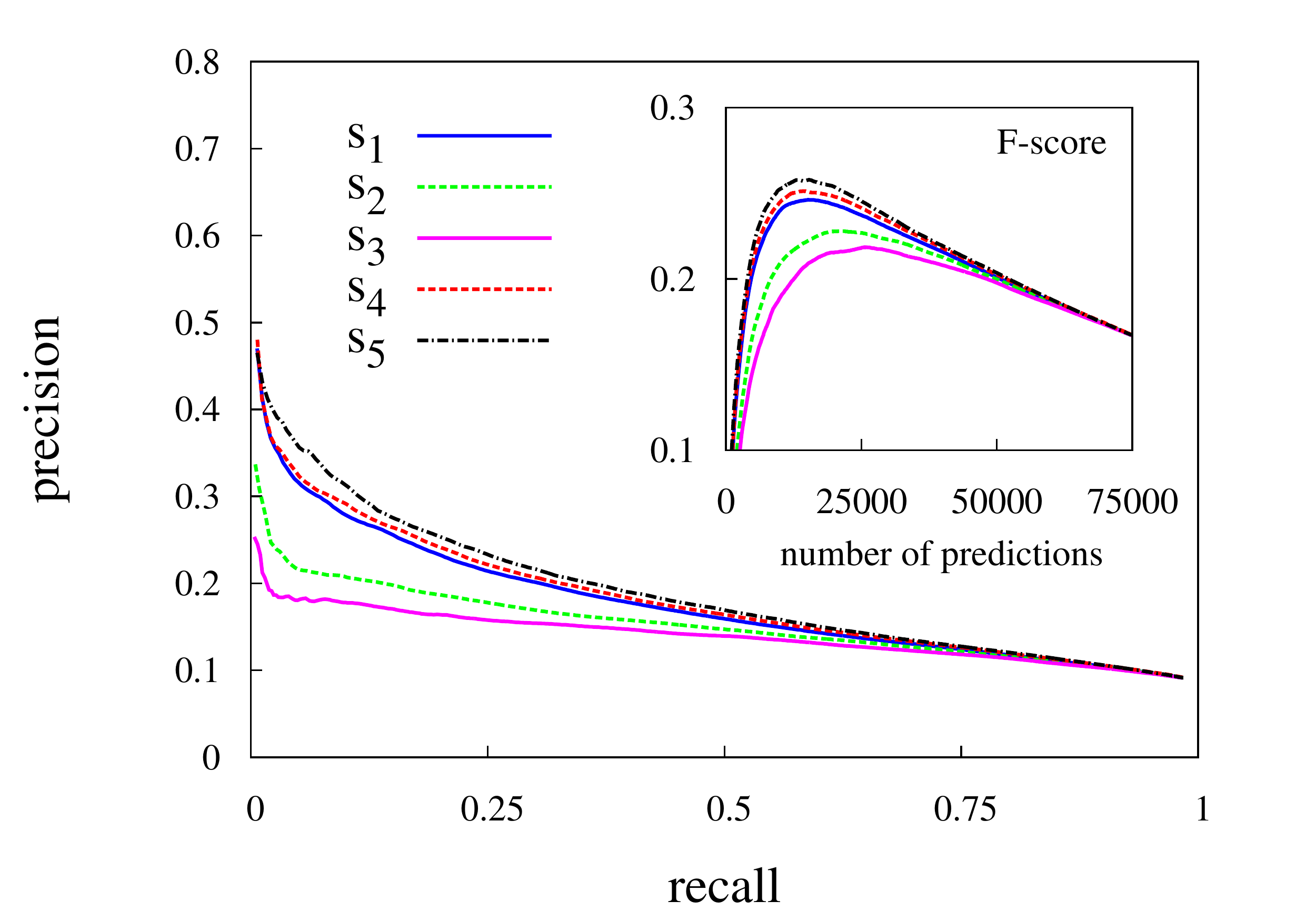}
\end{center}
\caption{\label{fig:pred_bench} Performance comparison between structural benchmarks, using precision vs recall and F-score (inset). Degree class: $ k \geq 10$ on the phonecall network, learning set.}
\end{figure}

Figure~\ref{fig:pred_bench} depicts the results of drawing randomly $ 1000 $ egos with $ k \geq 10$ from the learning set of the phonecall network.
It can be seen that $s_1$, $s_4$ and $s_5$ clearly outperform the two other metrics and the precision of $ s_5 $ is better for low recall predictions.
This observation stands using other samples and other classes.
Therefore, $ s_5 $ is used as the static benchmark of reference in the text that follows.

\subsection{Metrics using temporal information\label{sec:unsupervised}}

We aim at drawing as much information as possible from the temporal communication patterns of an ego to its neighborhood.
For this purpose we define weak classification metrics, which are complementary to each other as they use either different types of approaches or different timescales.

\subsubsection{Link strength metrics}

The first approach assumes that if there are strong links between $e$ and $i$, and $e$ and $j$, then $i$ and $j$ are more likely to be connected.
A straightforward way to measure the strength of a relationship is the total duration of phonecalls.
If $\Delta(e,i)$ is the total duration of phonecalls between $e$ and $i$, then we define the duration score as
$$ s_{dur}(i,j) = \frac{\Delta(e,i) \cdot \Delta(e,j)}{ \left(  \sum _k \Delta (e,k) \right) ^2}. $$

Strength may be measured in other ways, such as using the regularity of a relationship.
We can indeed expect that someone calls his or her relatives not necessarily often nor for a long time, but on a regular basis (every day or week for example).
We define the regularity $ \gamma (e,i) $ of a relationship as $w(e,i)$ divided by the Fano factor $F(e,i)$ of the inter-event time series. 
Let us recall that the Fano factor of a distribution is the ratio of its variance over its mean.
More regular signals are characterized by lower values of $F$  and, therefore, a  higher value of $ \gamma (e,i) $.
For $ \gamma (e,i) $ to be defined, we demand that there are at least two inter-event times in the time series (that is at least 3 interactions).
The regularity score is then defined as
$$ s_{reg} (i,j) =  \gamma (e,i) \cdot \gamma (e,j) .$$

\begin{figure}[h!]
\begin{center}
\includegraphics[angle=-90,width=0.6\linewidth]{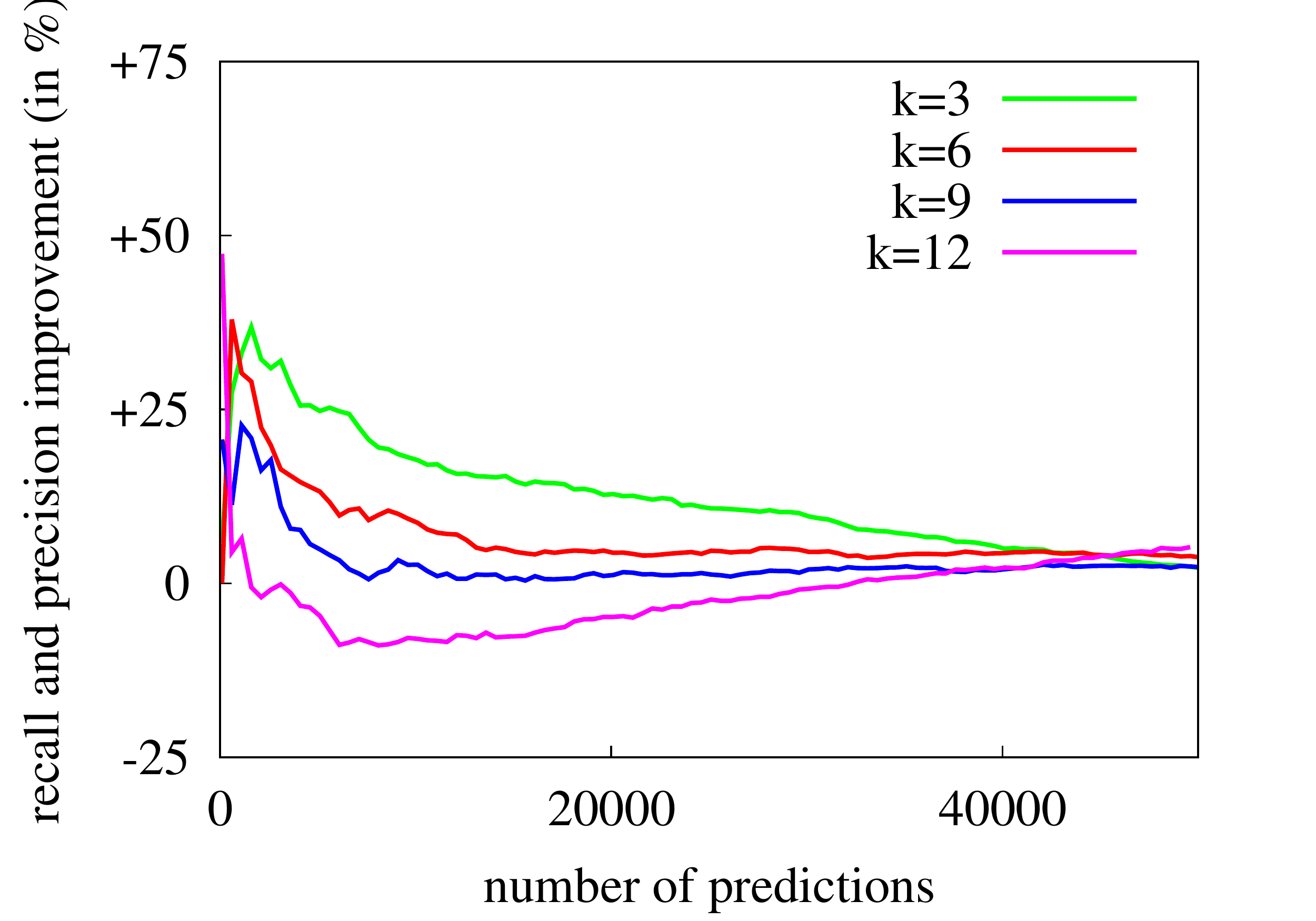}
\end{center}
\caption{\label{fig:duration_imp} 
Precision and recall improvements using the duration metric (on phonecalls, learning set) compared to $ s_5 $ benchmark for several degree classes.
}
\end{figure}

\begin{figure}[h!]
\begin{center}
\includegraphics[angle=-90,width=0.6\linewidth]{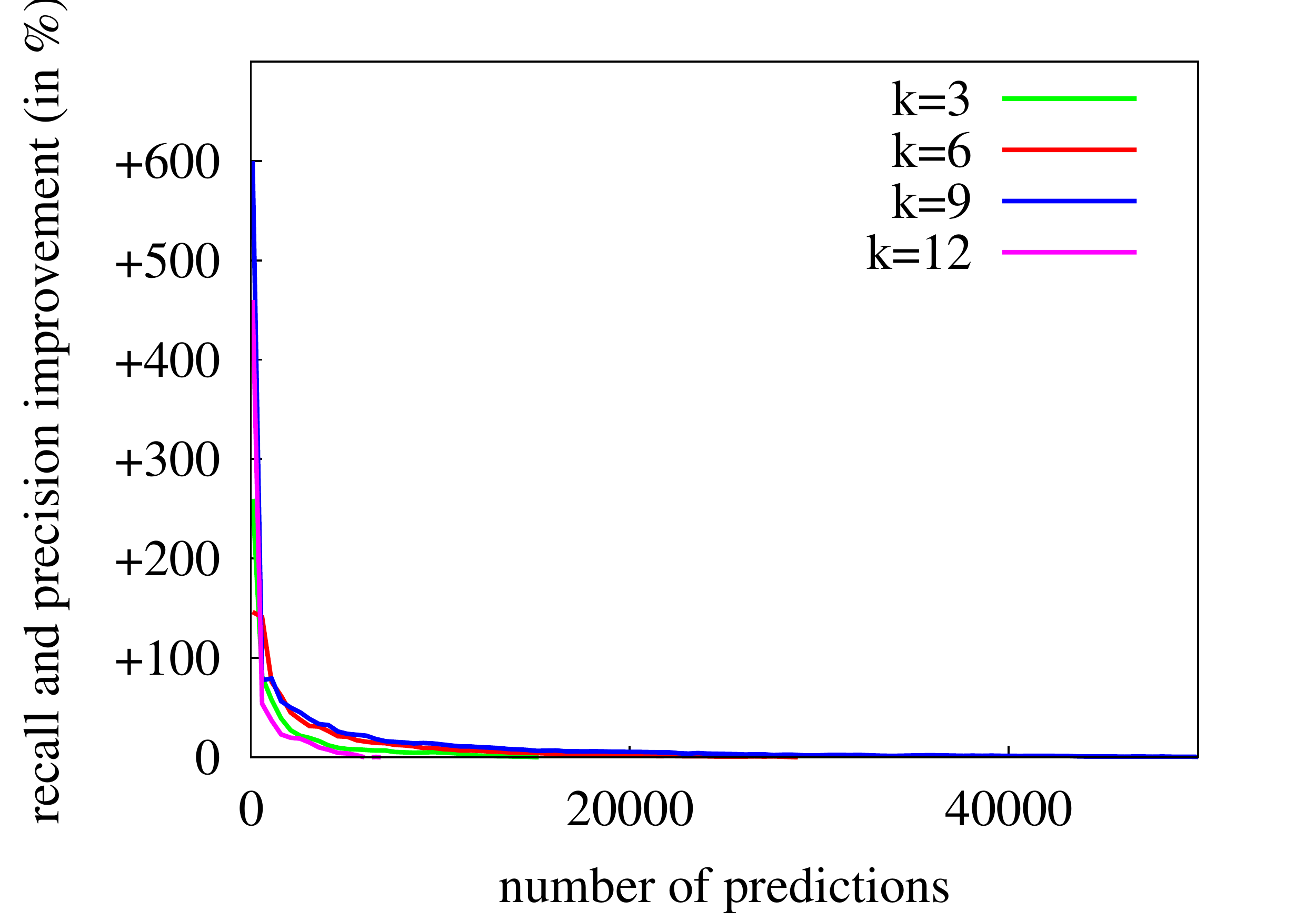}
\end{center}
\caption{\label{fig:regularity_imp} 
Precision and recall improvements using the regularity metric (on phonecalls, learning set) compared to $ s_5 $ benchmark for several degree classes.
}
\end{figure}

In Figures \ref{fig:duration_imp} and \ref{fig:regularity_imp}, we show the precision and recall improvements compared to the benchmark $s_5$, obtained respectively with the duration and regularity metrics.
Note that precision and recall improvements are equal for a fixed number of predictions.
Different degree classes are considered and it can be seen that there is an improvement to the benchmark in all cases except for $ k=12 $ with duration, where it is low or even negative.
In the case of the regularity metric, the improvement is spectacular for the first predictions but falls quickly to negligible values.
Considering duration, the improvement is not as high for the first few predictions but remains significant on a large range of predictions.

\subsubsection{Temporal profile approach}
\label{profilebased}

\begin{figure}[h!]
\begin{center}
\includegraphics[width=0.6\linewidth]{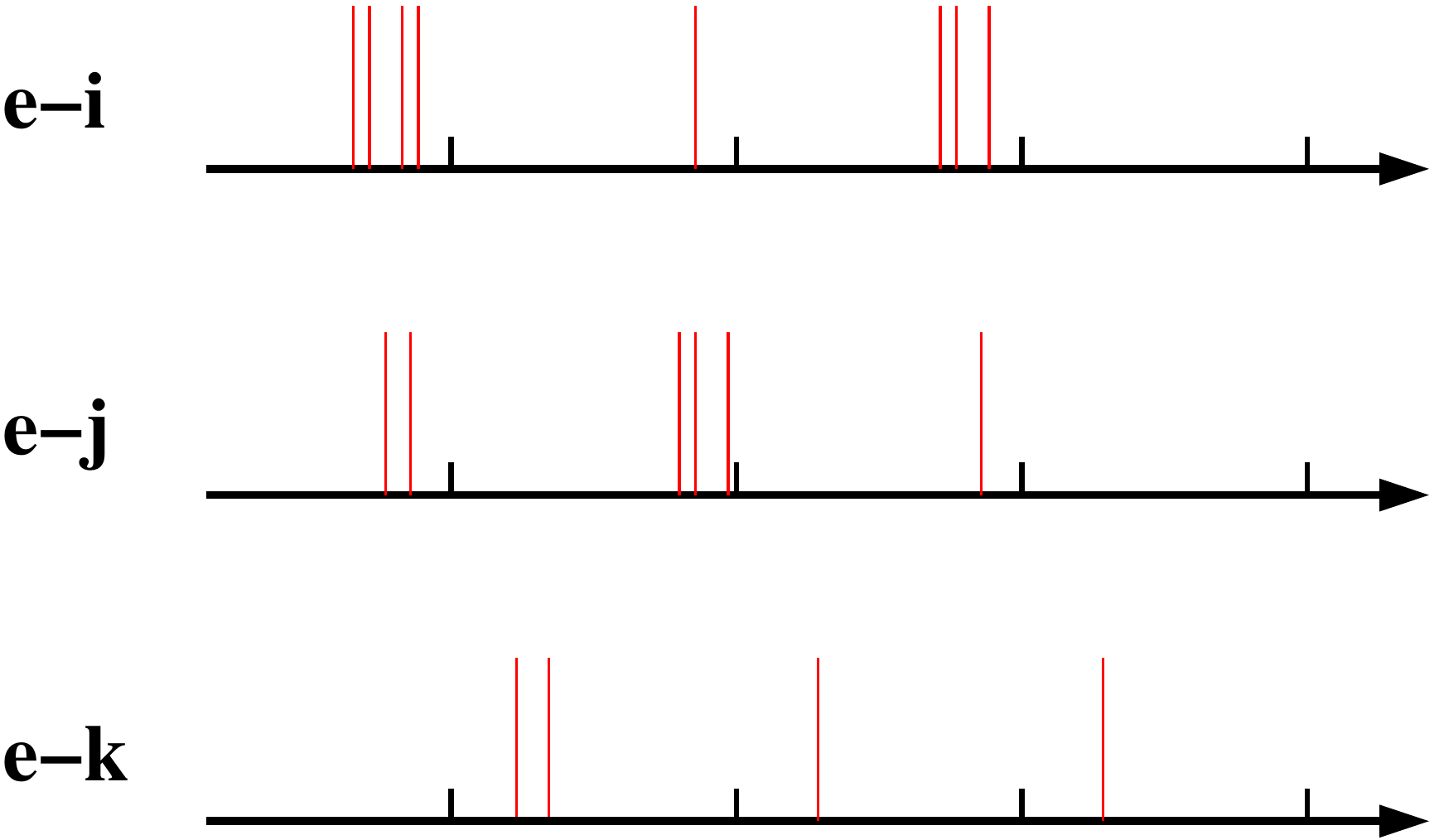}
\end{center}
\caption{\label{fig:ex_profiles} Example of temporal profiles of interactions between ego $e$ and neighbors $i$, $j$, $k$: a spike represents a phonecall.
The similarity of profiles indicates that $i$ and $j$ may be part of the same social circle, while $k$ is probably not.
}
\end{figure}

Depending on the moment of the day, week, or year, people use cellphones with different purposes.
For example, co-workers call each other more often during working days than during the week-end.
We, therefore, expect that the calling frequencies give clues about the underlying social groups.
This should reflect on temporal profiles as is shown in the example in Figure~\ref{fig:ex_profiles}.

We implement this idea in the following way. 
We divide the timeline $T$ in two sets of timestamps $T_A$ and $T_B$, and count the number of interactions during both periods by defining a $2$-dimensional weight vector, $( w_A(e,i) ; w_B(e,i) )$.
Assuming that pairs of nodes interacting with the central ego in a similar way are more prone to be connected, the score of the pair $ (i,j) $ is then computed from the scalar product of these weight vectors: 
$$s_{pr}(i,j) = \frac{(w_A(e,i) \cdot w_A(e,j) +  w_B(e,i) \cdot w_B(e,j))}{W(e)} .$$
Notice that $ s_{pr} = s_5 $ for $ T_A = T $ and $ T_B = \emptyset $.

We use the following profile scores in the rest of the study:
\begin{itemize}
\item[$ \bullet $] $ s_{pr-1} $ for a partition according to days of the week: Monday to Friday vs Saturday to Sunday,
\item[$ \bullet $] $ s_{pr-2} $ for a partition according to hours of the day: 8am to 6pm vs 6pm to 8am,
\item[$ \bullet $] $ s_{pr-3} $ for another partition according to hours of the day: 0am to 6pm vs 6pm to 0am,
\end{itemize}

\begin{figure}[h!]
\begin{center}
\includegraphics[angle=-90,width=0.6\linewidth]{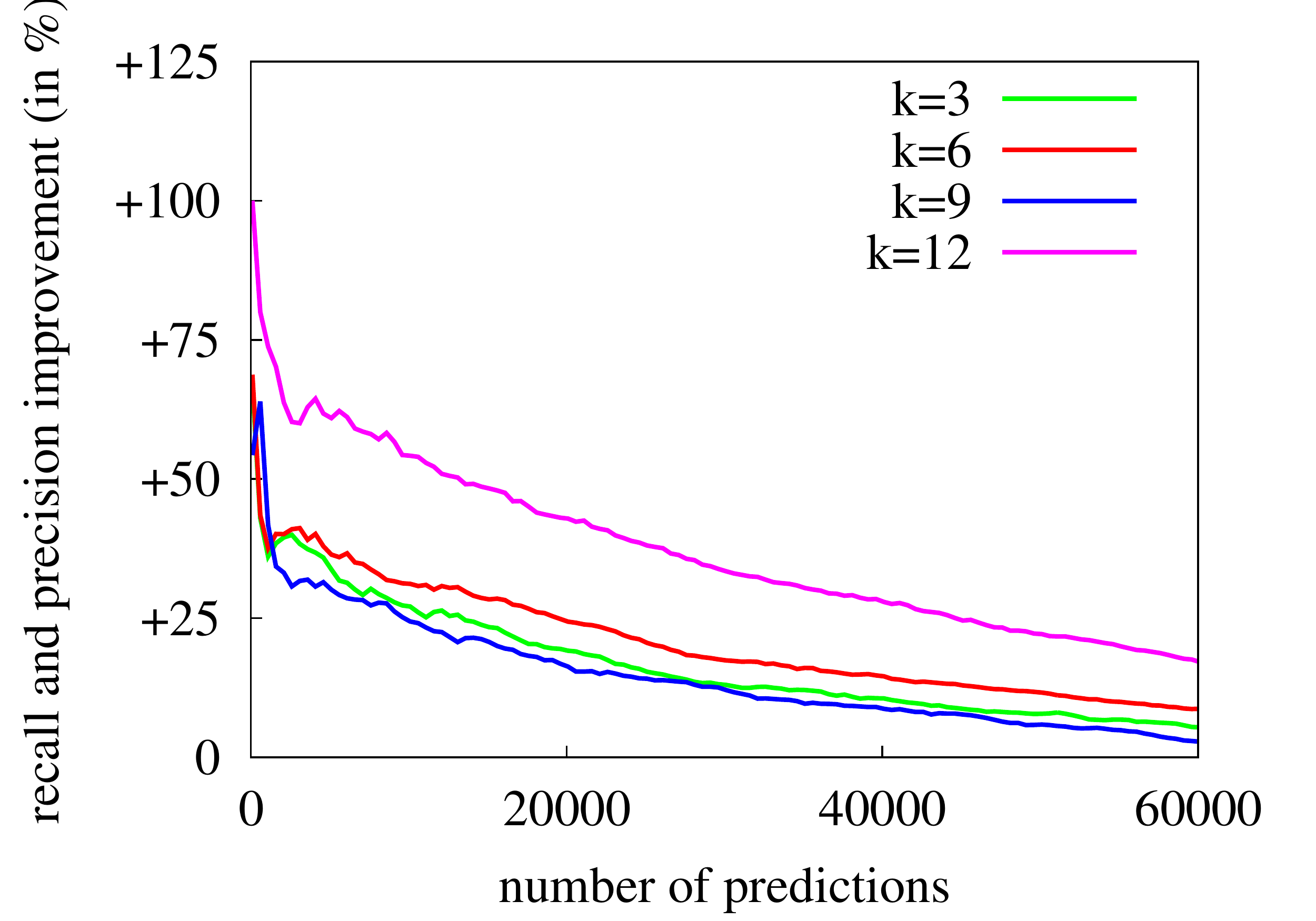}
\end{center}
\caption{\label{fig:temp_profile} 
Precision and recall improvements using the temporal profile approach (score $ s_{pr-1} $ on phonecalls, learning set) compared to $ s_5 $ benchmark for several degree classes.
}
\end{figure}

In Figure~\ref{fig:temp_profile} we summarize the precision and recall improvements compared to the benchmark $s_5$ obtained for different degree classes with profile $1$, where the timeline is partitioned between week days and week-end.
It reveals that $ s_{pr-1} $ performs much better than the benchmark, reaching up to a 67\%, 69\%, 66\% and 100\% enhancement for  classes $k=3$, $k=6$, $k=9$ and $k=12$ respectively.
Notice that the best improvements are obtained on the top-ranked pairs, which will be used in the aggregation we develop in Section~\ref{sec:supervised-merging}.

Of course, we can look for refined partitions of the timeline with more groups, more precisely defined boundaries, or even overlapping categories.
However, we take a different approach here by combining several weak classifying features to obtain a good ranking.\\

\subsubsection{Elapsed time approach}
\label{timebased}
%
When taking part in a social event, an individual has a high probability to call or to be called in a short period by several participants, for example, to set up a meeting point.
More generally, the elapsed time between calls may be an indication of a relationship between the users involved in both phonecalls.
That is why specific temporal patterns are found more often in phonecall networks than what is expected from randomized models (see~\cite{kovanen2011temporal,tabourier2012detect}). 
Such correlations appear at various timescales.
For example, defining a meeting point may involve several phonecalls within a few minutes, while the organization of a week-end may appear by examining patterns spreading over several hours or even days.

\begin{figure}[h!]
\begin{center}
\includegraphics[width=0.6\linewidth]{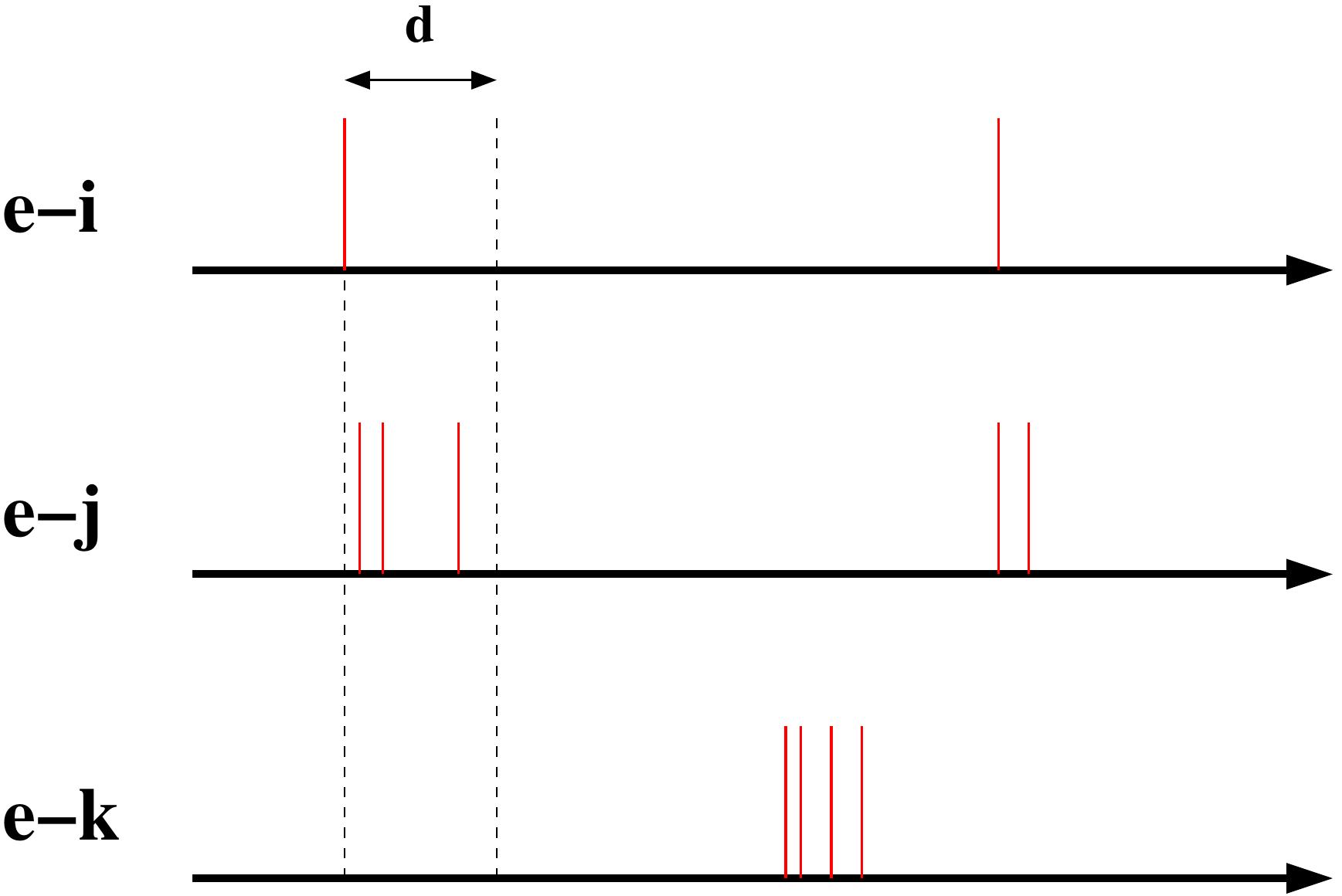}
\end{center}
\caption{\label{fig:time_elapsed_princ}
The time elapsed between $e-i$ and $e-j$ interactions is shorter than $d$, while it is not the case for $e-i$ and $e-k$.
We assume that it indicates a higher probability for $i$ and $j$ than for $i$ and $k$ to be part of the same social circle.
}
\end{figure}

In order to account for this mechanism,
we define a ranking score that takes into account the fact that an interaction between $i$ and $e$ took place not long before or after an interaction between $j$ and $e$. To do so, we define the pair score as a function of parameter $d$
$$ s_d(i,j) = \sum _{t_i,t_j} H[d - (t_j - t_i)] / W(e) ,$$
where $t_k$ is an interaction timestamp between $e$ and $k$, and $ H $ is the Heaviside function. 
In other words, each pair of interactions $(e-i , e-j)$ happening in a time shorter than $d$ increases the score of the pair $ (i,j) $.
This idea is represented schematically in Figure~\ref{fig:time_elapsed_princ}.
Note that $ s_{d=\infty} = s_5 $, as  $ \lim\limits_{d \to \infty} \sum _{t_i,t_j} H[d - (t_j - t_i)] = w(e,i) \cdot w(e,j)$.

\begin{figure}[h!]
\begin{center}
\includegraphics[angle=-90,width=0.6\linewidth]{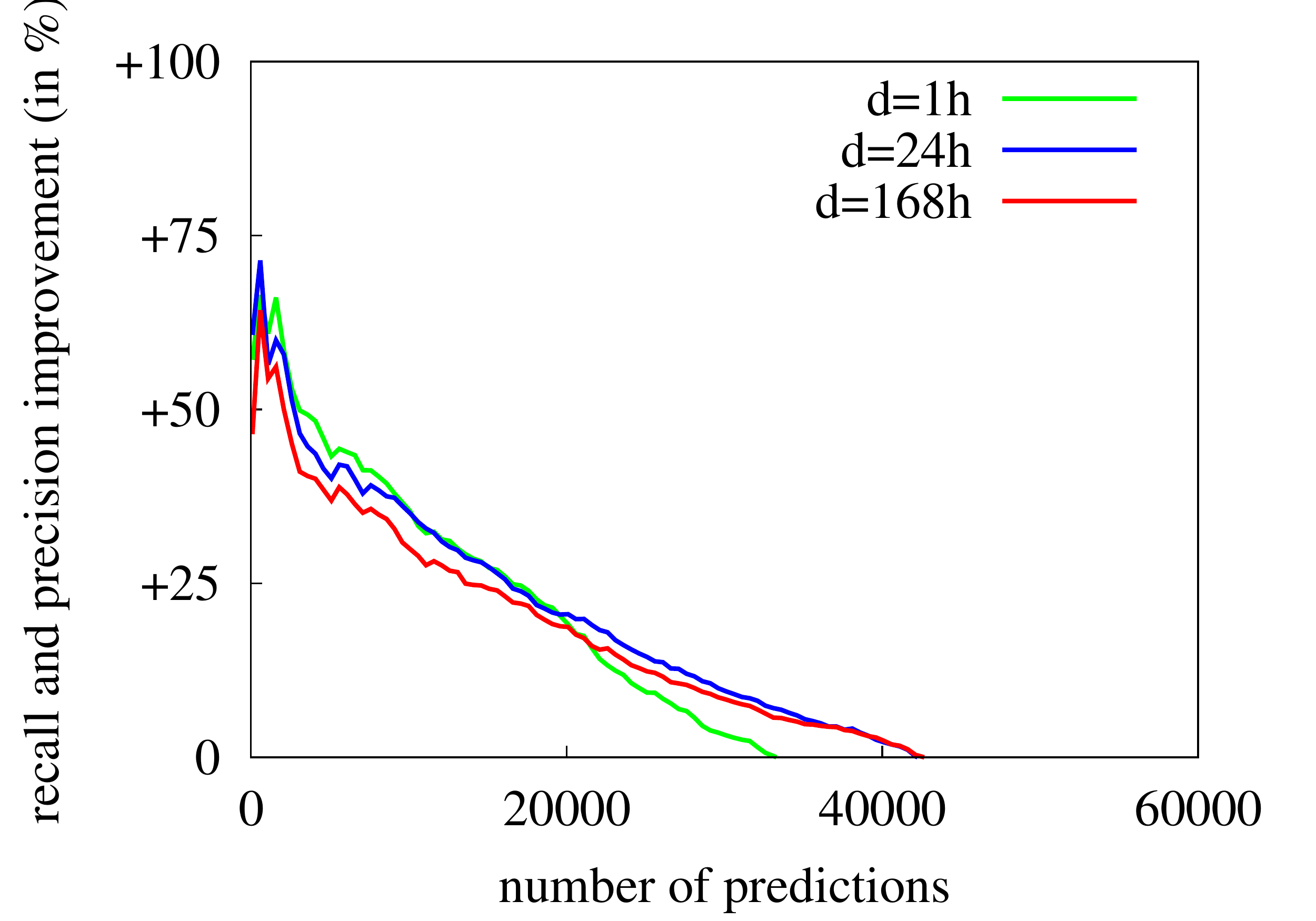}
\end{center}
\caption{\label{fig:time_elapsed_res} 
Precision and recall improvements using the time elapsed approach (on phonecalls, learning set) compared to $ s_5 $ benchmark for $d =$ 1 hour, 1 day or 1 week.
}
\end{figure}

Figure~\ref{fig:time_elapsed_res} shows results obtained for $ s_{d=1h} $, $ s_{d=24h} $ and $ s_{d=168h} $, corresponding respectively to a 1 hour, 1 day and 1 week time between phone calls for the degree class $ k=12$.
Here too, we see that there is a significant enhancement to the benchmark, and that the precision improvement curves are not equal for the different elapsed time parameters, meaning that different time scales bring different information. 
We will, therefore, combine the information brought by the various rankings to improve the quality of the predictions in the study that follows.

\section{Combining different predictors}
\label{sec:combin}

The ranking methods presented in the former section use temporal information in complementary ways. 
That is, we do not communicate in the same fashion with our family, friends, co-workers, etc.
Hence, a link detected as likely using a specific ranking method may not be discovered using another. 
In this section, we explore the possibility to combine the different rankings in order to obtain the best possible prediction.

\subsection{Feature selection and ranking correlations}

\begin{figure}[h!]
\begin{center}
\includegraphics[width=0.6\linewidth]{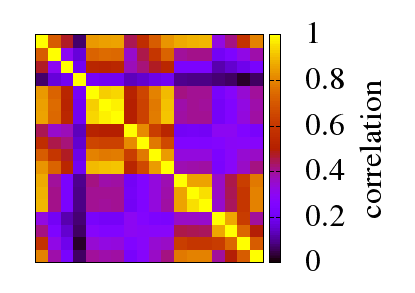}
\end{center}
\caption{\label{fig:corr_matrix}
Spearman correlation coefficients between rankings.
Ranking are ordered according to the following scores (left to right, up to bottom).\newline
Benchmark: $s_5$, duration based: $ s_{dur}^{phone} $, regularity based: $ s_{reg}^{phone} $, $ s_{reg}^{text} $, \newline
elapsed time based: $ s_{d=1h}^{phone} $, $ s_{d=3h}^{phone} $, $ s_{d=24h}^{phone} $, $ s_{d=168h}^{phone} $, $ s_{d=1h}^{text} $, $s_{d=3h}^{text} $, $ s_{d=24h}^{text} $, $ s_{d=168h}^{text} $,\newline
profile-based: $ s_{pr-1}^{phone} $, $ s_{pr-2}^{phone} $, $ s_{pr-3}^{phone} $, $ s_{pr-1}^{text} $, $ s_{pr-2}^{text} $, $ s_{pr-3}^{text}$.
}
\end{figure}

In the rest of our study, we use the 18 rankings corresponding to the following scores: $s_5$, $ s_{dur}^{phone} $, $ s_{reg}^{phone} $, $ s_{reg}^{text} $, $ s_{d=1h}^{phone} $, $ s_{d=3h}^{phone} $, $ s_{d=24h}^{phone} $, $ s_{d=168h}^{phone} $, $ s_{d=1h}^{text} $, $s_{d=3h}^{text} $, $ s_{d=24h}^{text} $, $ s_{d=168h}^{text} $, $ s_{pr-1}^{phone} $, $ s_{pr-2}^{phone} $, $ s_{pr-3}^{phone} $, $ s_{pr-1}^{text} $, $ s_{pr-2}^{text} $, $ s_{pr-3}^{text}$.
To support the idea that different rankings bring different information, we measure the correlation between these 18 rankings and represent in Figure~\ref{fig:corr_matrix} the Spearman correlation coefficient matrix between rankings in the case of degree class $ k=12 $.
Correlations for other degree classes look similar, but are not reported here for the sake of brevity.
We observe that correlations are heterogeneous.
For example $ s_{reg}^{text} $ is lowly correlated to all other rankings, whereas $ s_5 $ is quite highly correlated to a majority of rankings.
Groups of metrics can be distinguished based on the correlation matrix, while $ s_{dur}^{phone} $, $ s_{reg}^{phone} $ and $ s_{reg}^{text} $ are relatively independent from the others.
The profile-based classifiers of Section \ref{profilebased} are on average highly correlated and the same can be said for the elapsed time-based classifiers of Section \ref{timebased}, as is expected.
We also notice that these two groups can be divided into two subgroups, corresponding respectively to phone and text-messages classifiers.

On the whole, it appears that some pairs of nodes are ranked high according to a classifier, but not by all others.
In the following study, we present ways to draw benefit from the complementarity of these scores.

\subsection{Unsupervised consensus methods}

We describe here unsupervised techniques used to merge rankings based on social choice theory~\cite{dwork2001rank}.
These methods are consensus-based.
They rely on the assumption that every ranking provides a reasonable solution to the problem and combine rankings by giving to each of them an equal weight.

\subsubsection{Borda's method}

Borda's method is a \textit{rank-then-combine} method, originally proposed to obtain a consensus in a voting system~\cite{deborda1781memoire}.
We use the index $ \kappa $ to refer to a specific ranking among the $\alpha$ rankings combined. 
Hence, $ r_\kappa (i,j) $ denotes the rank of pair $ (i,j) $ according to this ranking, and $ | r_\kappa |$ denotes the number of elements ranked in $ r_\kappa $.
Each pair is given a score corresponding to the sum of the number of pairs ranked below, that is to say
$$ s_B (i,j) = \sum \limits_{\kappa = 1} ^{\alpha} | r_\kappa | -  r_\kappa (i,j)  .$$

This scoring system may be biased by the fact that some rankings feature less elements than others. 
To alleviate this problem, unranked pairs in ranking $ r_\kappa $, but ranked in $ r_{\kappa'} $ will be considered as ranked in $ r_\kappa $ on an equal footing as any other unranked pair, and below all ranked pairs of $ r_\kappa $.
Borda's method is computationally cheap (linear in the ranking size), which is a highly desirable property in our case, where many items are ranked.
A comprehensive discussion of this method can be found in~\cite{dwork2001rank}.

\subsubsection{Medrank}

Borda can be described as building the ranking by averaging the rankings combined.
Another possibility is to look for the median of the rankings.
The output, that is to say the combined ranking, is initially empty and built iteratively in the following way.
At step $n$ of the algorithm, the user register which pairs are ranked in position $n$ of every ranking and how many times each pair has been seen until then.
As soon as a pair $(i,j)$ has been seen in half (or more) of the number of rankings it belongs to, it is appended to the list representing the combined ranking.
Going through all rankings from top to bottom simultaneously, we obtain a ranking which can be interpreted as the median ranking of the input rankings.
This consensus method is called Medrank~\cite{sculley2007rank}, and it is also linear in terms of computational complexity.


\subsection{Classical Supervised classification methods \label{sec:supervised-classif}}

Another class of merging techniques proceeds in a supervised way.
Let us first introduce traditional classification methods. 
The rankings obtained with unsupervised methods on the learning set are the scores used as input features.
Then the link prediction issue is considered as a two-classes classification problem: the model trained on the learning set is applied on the test set to estimate if a link does exist or not.

For this purpose we used three different methods: \textit{Classification Trees}, \textit{AdaBoost} and \textit{k-Nearest Neighbors}, as documented in the python toolkit scikit-learn\footnote{\url{http://scikit-learn.org/}}.
One of the drawbacks of these methods is that the operator cannot set the number of predictions.
We, therefore, explore a small part of the precision-recall space using the parameters of the method.

\begin{figure}[!h]
\begin{center}
\includegraphics[angle=-90,width=0.45\linewidth]{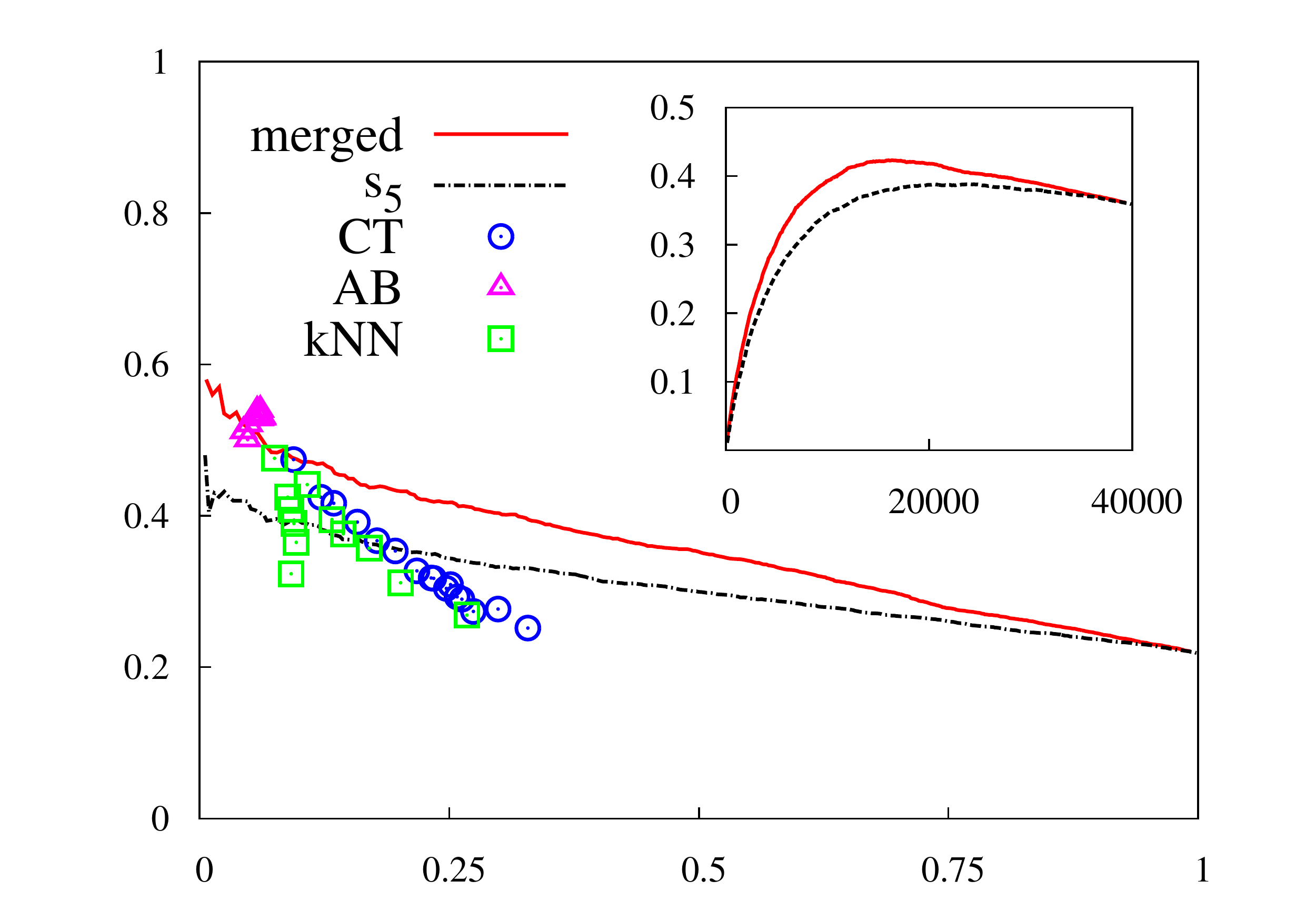}
\includegraphics[angle=-90,width=0.45\linewidth]{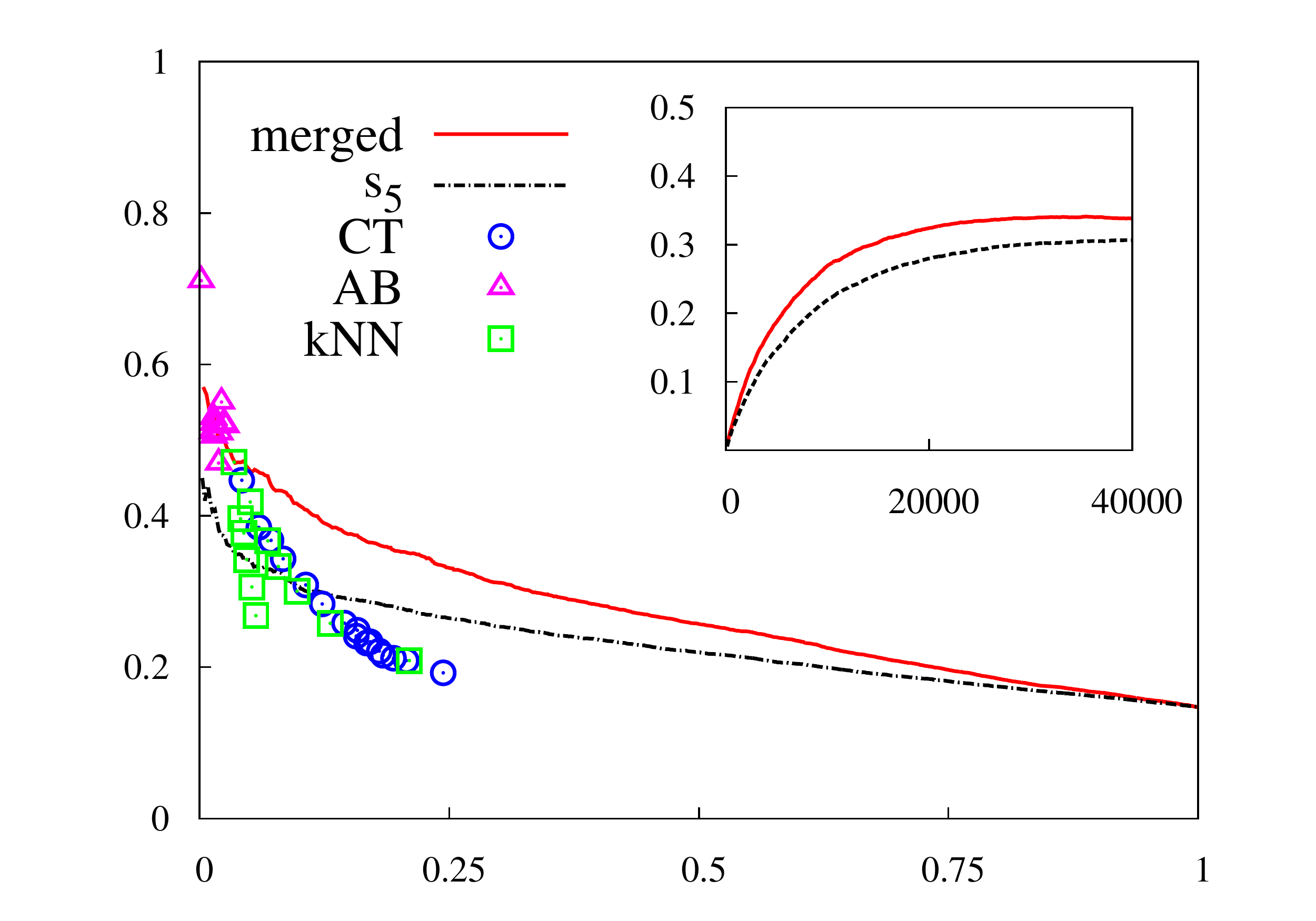}
\includegraphics[angle=-90,width=0.45\linewidth]{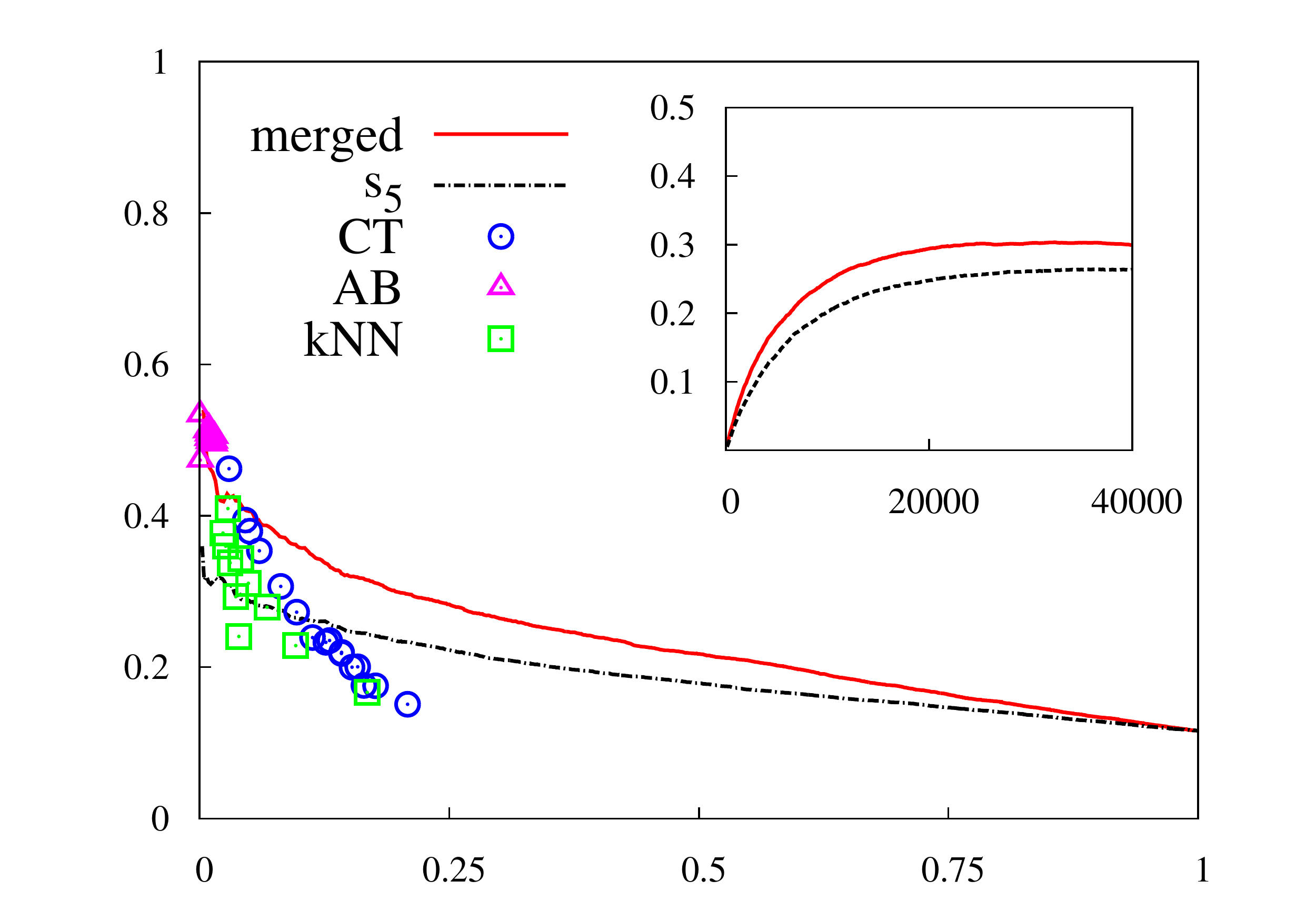}
\includegraphics[angle=-90,width=0.45\linewidth]{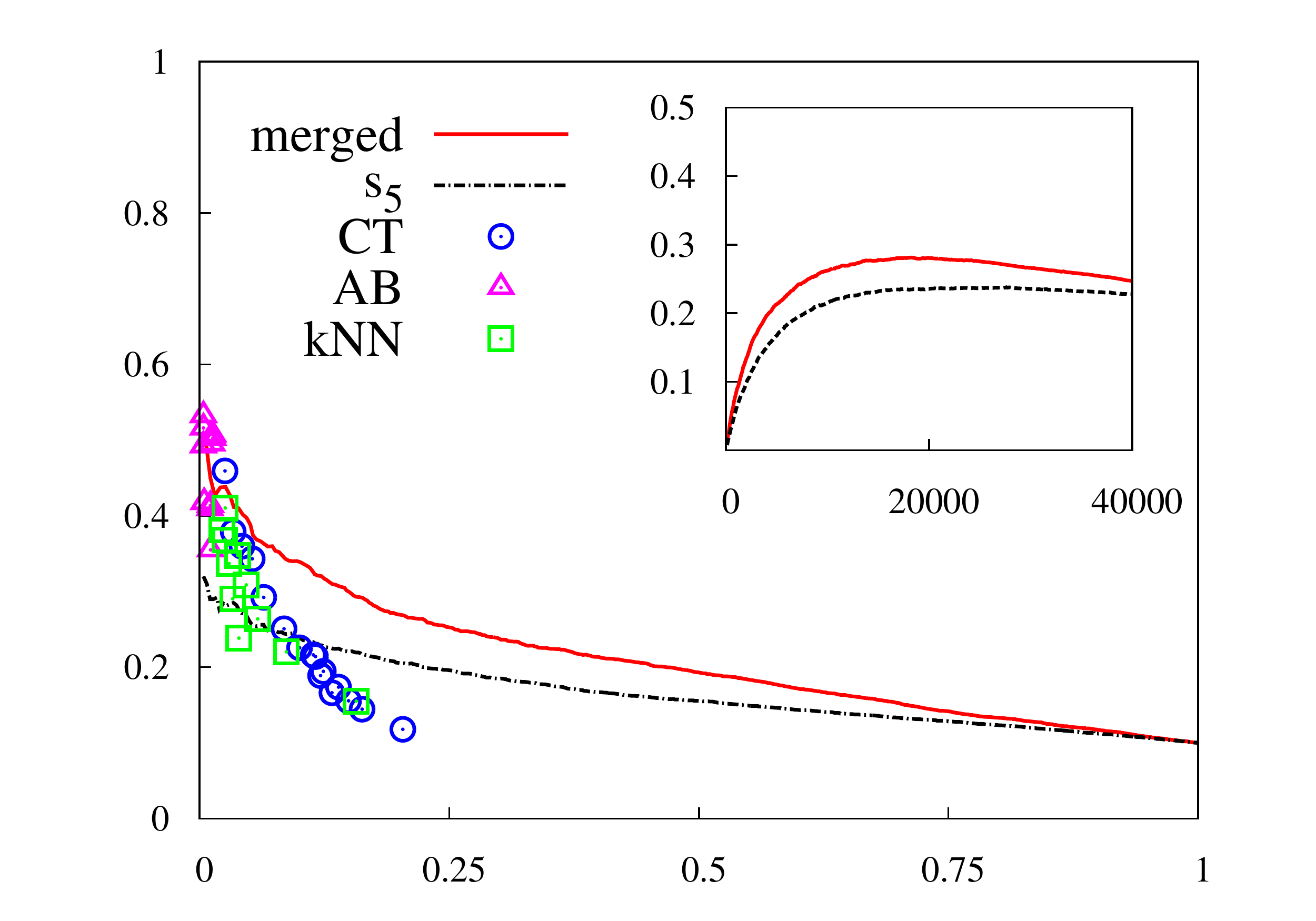}
\includegraphics[angle=-90,width=0.45\linewidth]{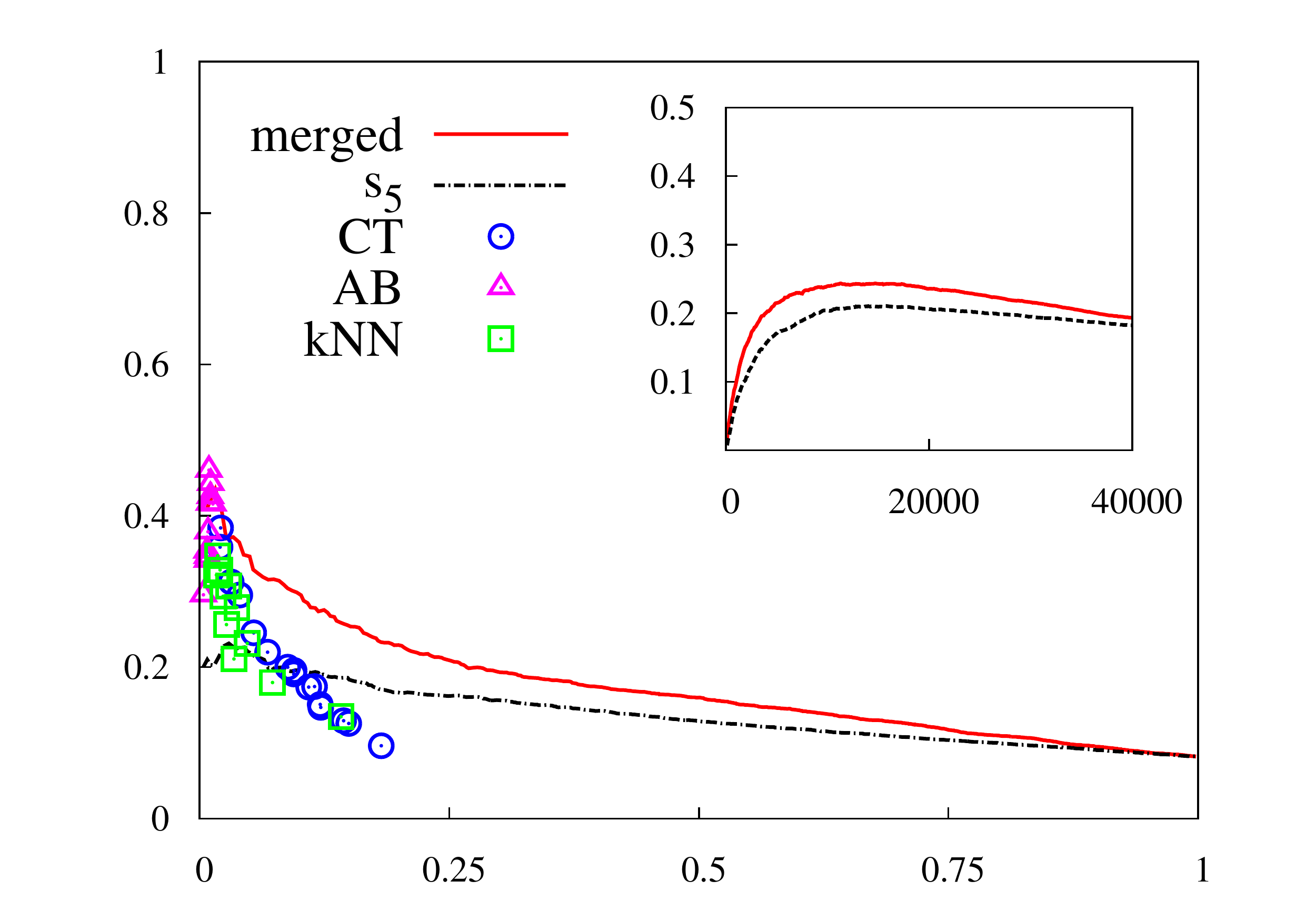}
\includegraphics[angle=-90,width=0.45\linewidth]{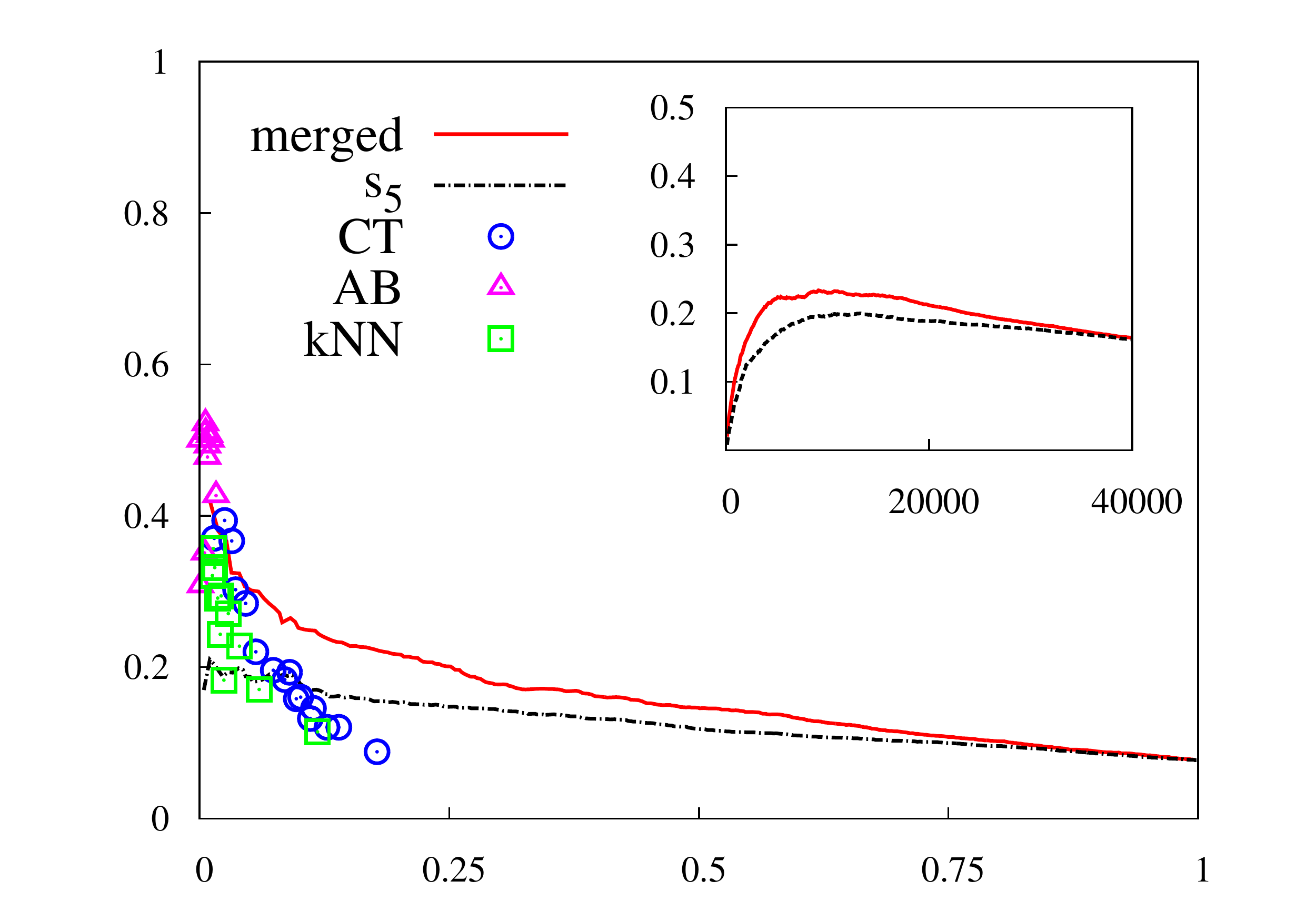}
\caption{\label{fig:res_per_class}
Left to right, top to bottom: $k=2$, $k=4$, $k=6$, $k=8$, $k=10$, $k=12$.
Main plots: precision vs recall.
Insets: F~scores against the number of predictions.
Blue circles correspond to Classification Tree (CT) results, purple triangles to AdaBoost (AB) and  green squares to k Nearest Neighbors (kNN).
}
\end{center}
\end{figure}

The results obtained are displayed in Figure~\ref{fig:res_per_class}, they show that these methods are efficient  to make high precision and low recall predictions (especially \textit{AdaBoost}), clearly outperforming the static benchmark $ s_5 $.
They are nonetheless inappropriate to make effective predictions over a large range of the precision-recall space.

\subsection{Supervised learning-to-rank methods \label{sec:supervised-merging}}

Finally, we use \textit{RankMerging}, a supervised machine learning framework \cite{tabourier2014rankmerging}, recently developed to aggregate information from various ranking techniques, in a way that is suited to link prediction.
Here we do not describe the algorithm in details and only focus on the points which are important for this study.
Notice that other learning-to-rank techniques could be used following the same scheme, but our framework is built for such situations with many ranked items as it is computationally linear.
Moreover, it does not demand for a pair to be highly ranked according to all criteria, but at least one, which we believe is appropriate in the context of link prediction in a social network.
Finally, it allows to investigate which features contribute to the final ranking and thereby, giving indications about the information sources which are important.

According to our framework, we first create the 20 rankings defined in the former parts (18 unsupervised score-based rankings plus Borda and Medrank) on each of the three sets (learning, validation and test).
Then we evaluate during the learning phase the coefficients that compute the contribution of each of these rankings to the merged ranking on the labelled learning set to optimize the quality of the prediction.
In more details, to create a combined ranking of length $n$, we learn the fraction $ \phi_\kappa $ of pairs which are extracted from each input ranking $ r_\kappa $.
$ \phi_\kappa $ coefficients are computed to maximize the quality of the prediction on the learning set.
The closer $ \phi_\kappa $ is to $1$, the heavier the weight of ranking $ r_\kappa $ in the merging process.
The only parameter of the method (called $g$ in~\cite{tabourier2014rankmerging}) is fixed on the cross-validation set to get the best prediction quality.
Finally, the performance of the whole process is evaluated by measuring the improvement of the prediction on the test set, compared to the static benchmark defined in Section~\ref{sec:bench}.
The performance will be measured using the area under the curve in the precision-recall space.


Results of \textit{RankMerging} on the test set are displayed in Figure~\ref{fig:res_per_class} and Table~\ref{tab:res}, degree class per degree class.
In general, predictions are more accurate for low-degree than for high-degree classes, which is a consequence of the fact that the clustering coefficient in a phonecall network is higher for low-degree nodes \cite{onnela2007analysis}.
Hence, it should be easier to target connected pairs.
However, the improvement to $s_5$ benchmark is higher for high degree-classes. 
It is well-known that the higher the degree of an ego, the higher its activity~\cite{miritello2013time}, so that we have access to a richer temporal information on high-degree ego networks to improve the predictions. 

\begin{table}[h!]

\caption{Improvement to benchmark $ s_5 $ of the area under the curve in the precision-recall space.
\label{tab:res}}

\newcolumntype{M}[1]{>{\centering}m{#1}}

\begin{center}
{
\begin{tabular}{M{2cm}|c} 
Ego degree  & Pr-Rc   \\
class & improvement  \\
\hline
$k=2$ & + 15.5\% \\
$k=3$ & + 18.8\%  \\
$k=4$ & + 19.3\% \\
$k=5$ & + 21.4\%\\
$k=6$ & + 22.3\% \\
$k=7$ & + 22.5\% \\
$k=8$ & + 25.5\% \\
$k=9$ & + 25.5\% \\
$k=10$ & + 28.1\%  \\
$k=11$ & + 30.9\% \\
$k=12$ & + 26.4\%  \\
$k=13$ & + 33.1\% \\
$k=14$ & + 36.2\%  \\
$k\geq15$ & + 51.6\%  \\
\end{tabular}
}
\end{center}
\end{table}

\subsection{Contribution of rankings and discussion}

\begin{figure}[!h]
\begin{center}
\includegraphics[angle=-90,width=0.6\linewidth]{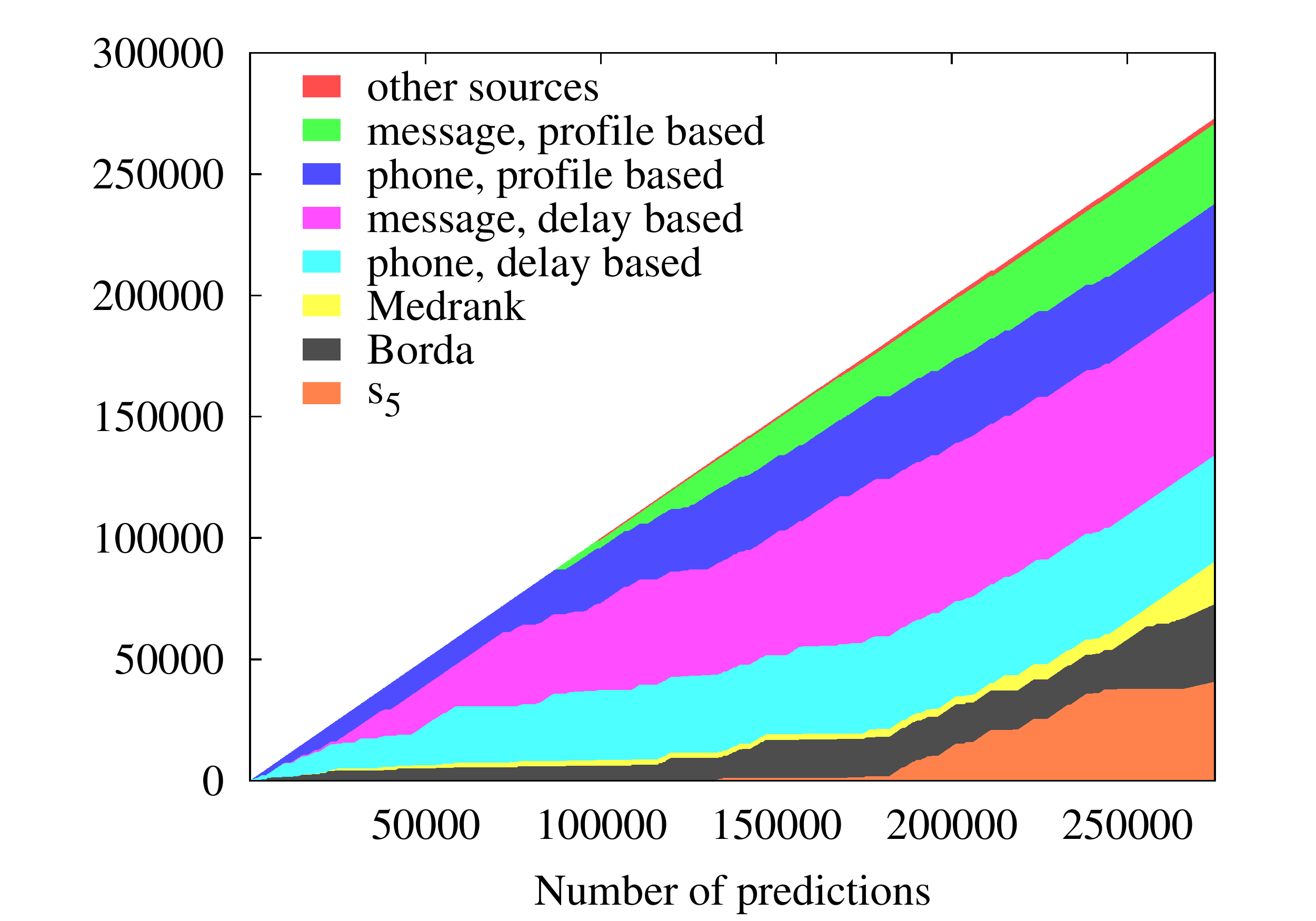}
\caption{\label{fig:weights_ranking}
Contributions of each ranking to the merged ranking, class $k=8$.
}
\end{center}
\end{figure}

We want to measure the contribution of each ranking to the merged ranking in order to evaluate its weight in the aggregation process.
\textit{RankMerging} allows to do so by indicating how many pairs of each ranking has been taken into account to create the merged ranking (for more details see~\cite{tabourier2014rankmerging}).
A number of pairs close to the number of predictions therefore indicates that a ranking has a heavy weight in the merging process.
We show in Figure~\ref{fig:weights_ranking} the contribution of each group of rankings to the process in the case of degree class $ k=8 $.

\begin{figure}[!h]
\begin{center}
\includegraphics[angle=-90,width=0.45\linewidth]{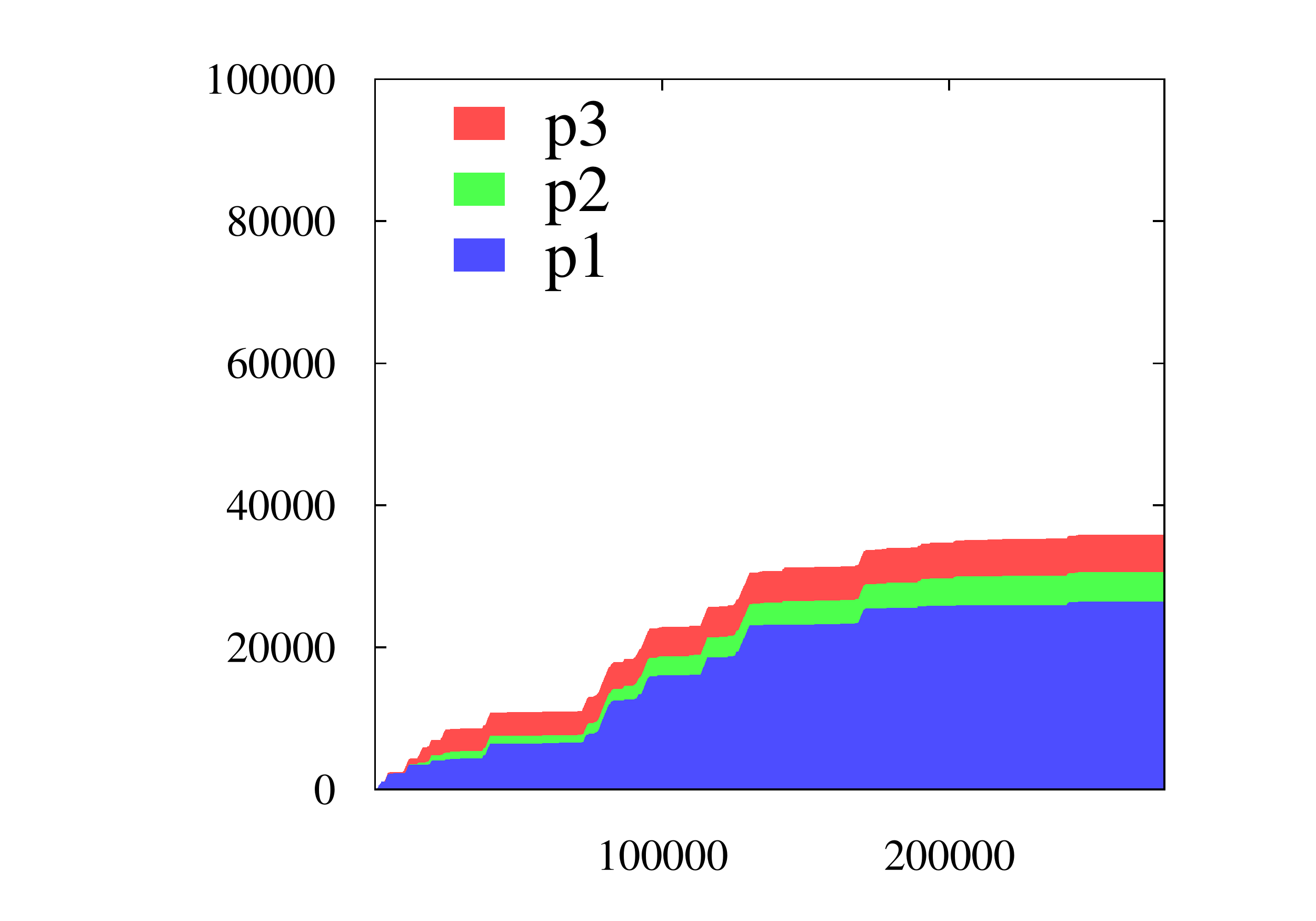}
\includegraphics[angle=-90,width=0.45\linewidth]{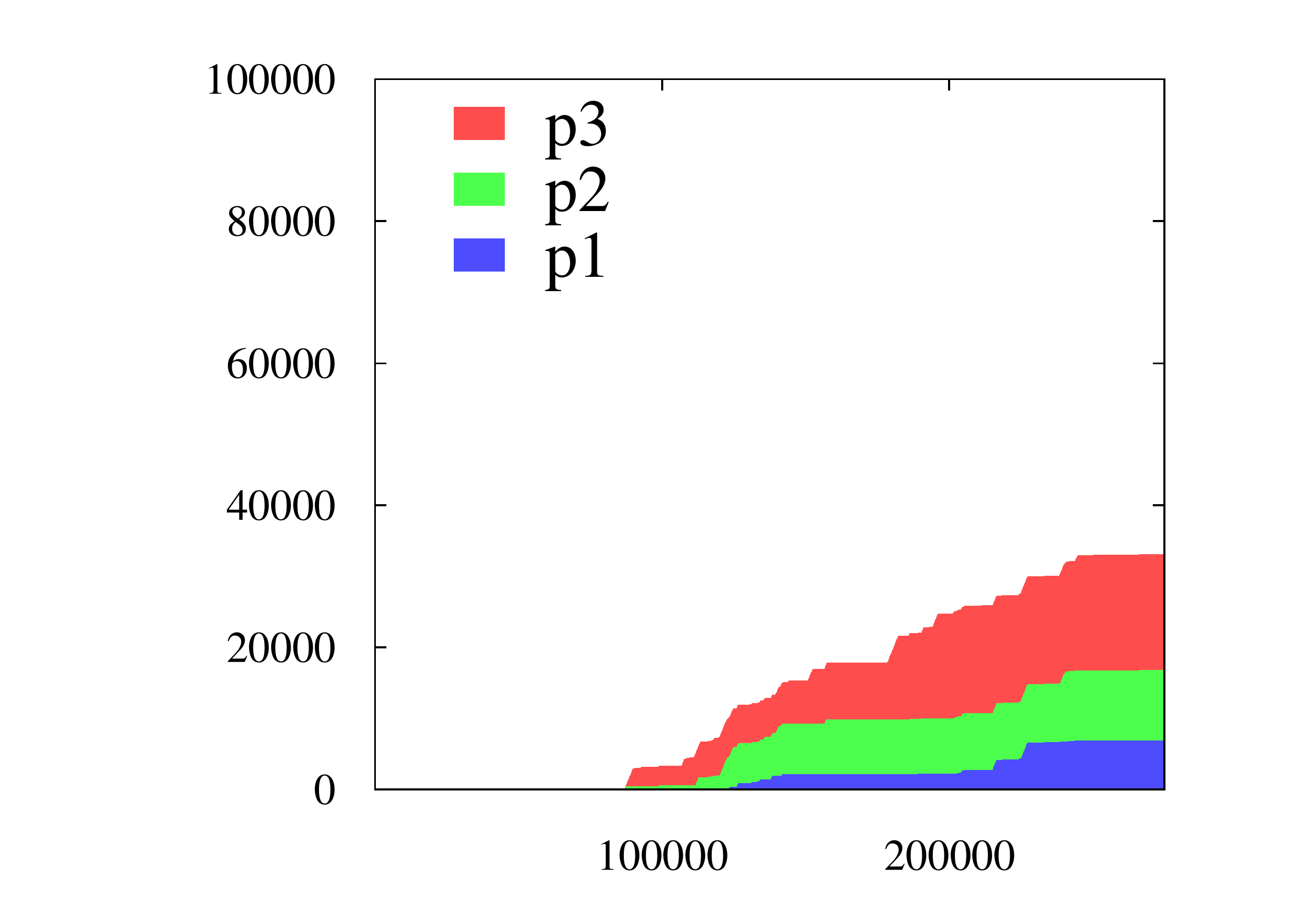}
\includegraphics[angle=-90,width=0.45\linewidth]{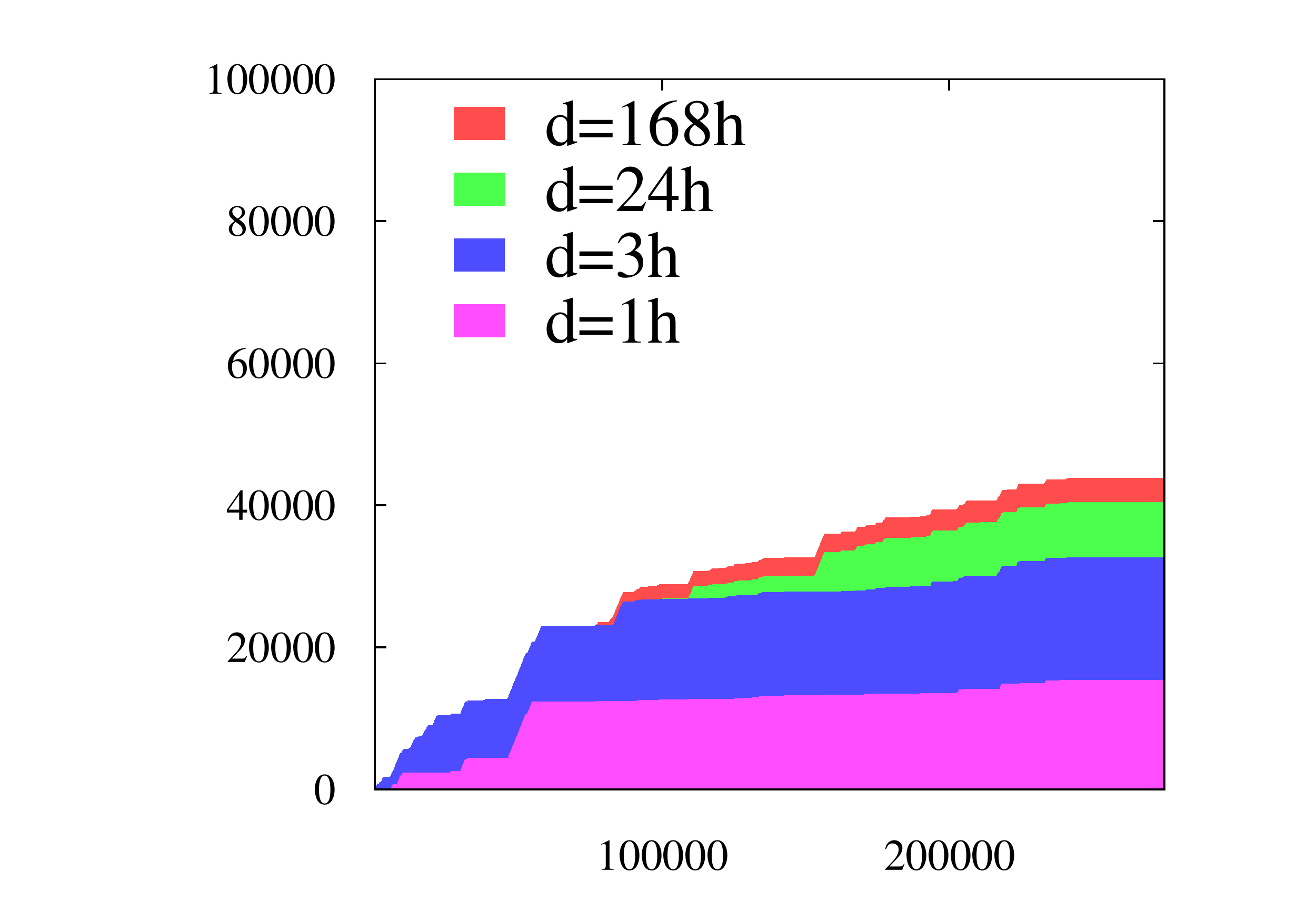}
\includegraphics[angle=-90,width=0.45\linewidth]{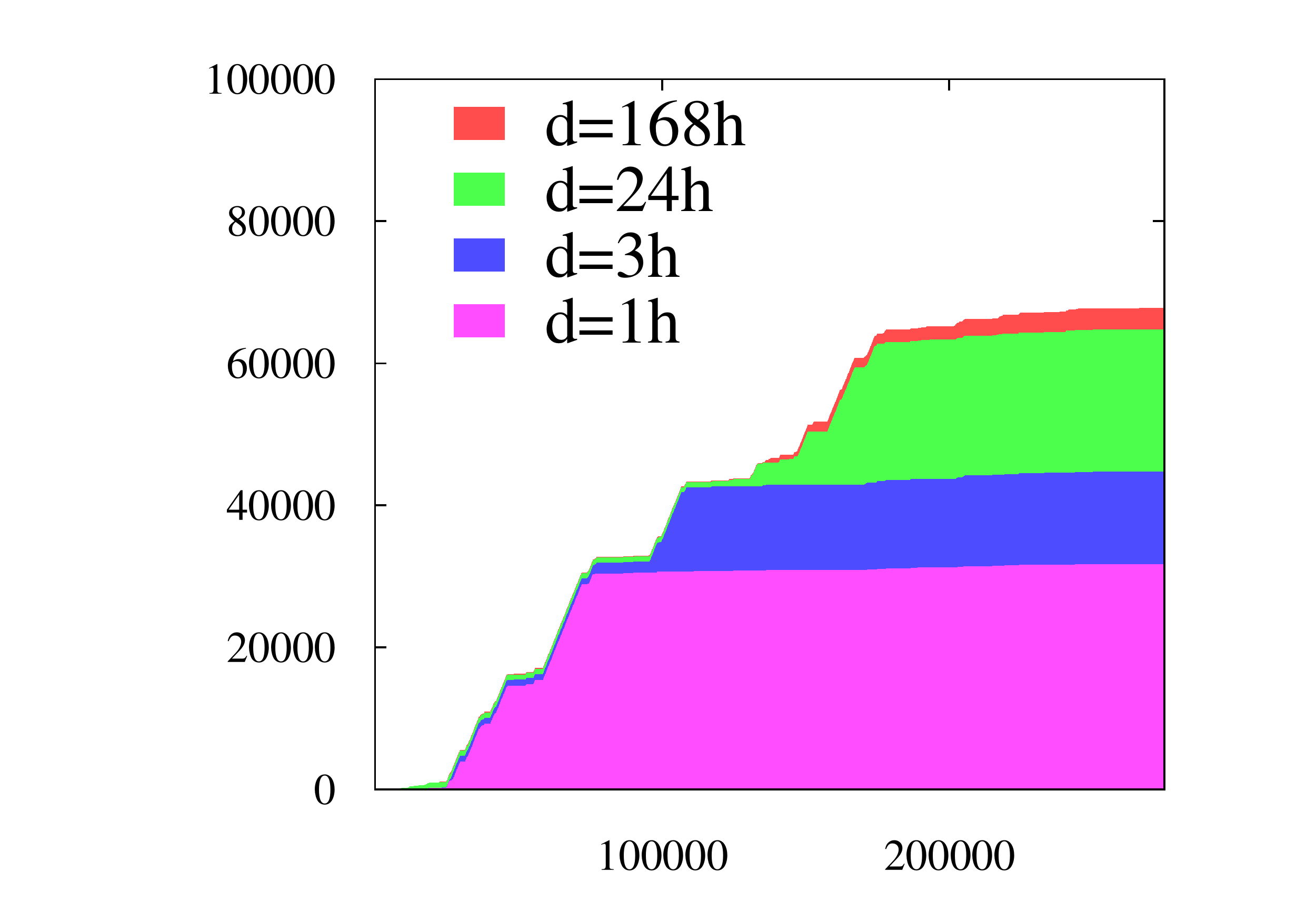}
\caption{\label{fig:weights_ranking_detail}
Contributions of each ranking to the merged ranking.
Top left: phonecalls profile-based,
top right: text messages profile-based,
bottom left: phonecalls elpased time-based,
bottom right: text messages elapsed time-based.
}
\end{center}
\end{figure}

We display a refined analysis on the elapsed time-based and profile-based predictions in Figure~\ref{fig:weights_ranking_detail} to have an idea of the contribution of each profile (resp. elapsed time) within each category.

Several trends can be seen in these graphs as mentioned below.
\begin{itemize}
\item Some classifiers are very little explored or even not used at all during the process probably because the information that they bring is redundant with other classifiers. 
This is the case of $ s_{reg}^{phone}$, $ s_{reg}^{text}$, and $ s_{dur} $ on this specific example.
\item During the first steps the rankings used are mostly profile-based and elapsed time-based.
As the first predictions correspond to the highest scores, these steps correspond to high precision and low recall, that is the top-ranked items of the merged ranking.
It means that these two sets of features may be considered as informative time-based predictors on this dataset.
\item A more thorough observation reveals that the information brought by phone calls is used more for the first predictions while text messages are used later in the process.
More precisely, the most used ranking during the first steps is related to the phonecall, elapsed time-based score.
\item Borda and Medrank are used during the whole process, which could be expected as these methods are designed to be an average or a median of all the others. In our case it seems that Borda's aggregation is much more informative than Medrank.
\end{itemize}

Notice that the class $ k=8 $ was taken as an example of a typical behaviour.
There are quantitative variations from a class to another.
However, the trends identified previously remain true with the other classes.

The fact that a ranking is used early in the process tends to prove that the information that it brings is relevant for link prediction.
From this observation, we suggest several conclusions related to the social meaning of our experiments. 
First, the most efficient classifier is the time elapsed between interactions, especially phone calls.
It seems indeed that calls separated by less than a few hours have a significant probability to involve members of the same social circle.
On the other hand, regularity-based classifiers proved themselves inefficient when aggregated.
Very regular interactions are probably too rare to allow the identification of a large number of social circles where it is a standard communication pattern. 
The duration based classifier brings little improvement to the prediction too.
However, the cause may be different as duration score is quite highly correlated to other scores, while regularity is not.
We suggest that duration is ignored during the combination process because it brings redundant information.
Finally, profile-based predictors appear as moderately efficient.
But interestingly, they seem complementary with the elapsed time-based predictors.
A possible interpretation is that there are social circles where people call or send messages according to a certain schedule, and others where interactions are rather triggered by other interactions.
This conclusion is of course hypothetical and calls for additional investigation.

\section{Conclusion}

In this article, we explored how it is possible to infer links in ego-networks, where the only information available is the timing of interactions of ego to its neighbors.
We proposed several ways of extracting information from the temporal communication patterns and showed that they can largely improve predictions when compared to a prediction based on the static information available - that is to say the weights of interactions.
More precisely, it seems that profiling interactions based on when ego communicates with other users and measuring the elapsed time between interactions are two particularly efficient techniques to infer which of ego's neighbors are likely to interact.
Our study also supports that depending on the kind of social relationship, communication modes vary, as we observed that different features as well as different time-scales reveal different links.
We took advantage of this for link prediction by using a learning-to-rank framework that may rank high items even if some features do not rank them high.

We studied a case, where structural information is minimal and therefore, isolating how the temporal features that we defined improved the prediction.
However, this temporal-based approach can be advantageous even if we have richer information on the network since it provides additional sources of information for link inference.
It could for example be used to predict future interactions.
Knowing the current state of a social network as well as the dynamics of existing interactions, it would improve our knowledge of the active social circles and potential new interactions.


\section*{Acknowledgements}

This paper presents research results of the Belgian Network DYSCO (Dynamical Systems, Control, and Optimization), funded by the Interuniversity Attraction Poles Programme, initiated by the Belgian State, Science Policy Office. The scientific responsibility rests with its authors.
Computational ressources have been provided by the Consortium des \'Equipements de Calcul Intensif (C\'ECI), funded by the Fonds de la Recherche Scientifique de Belgique (F.R.S.-FNRS) under Grant No. 2.5020.11


\bibliographystyle{bmc-mathphys} 
\bibliography{biblio}  


\end{document}